\documentclass[12pt]{elsarticle}

\usepackage{hyperref}
\usepackage[utf8]{inputenc}
\usepackage{color}
\usepackage{amsmath}
\usepackage{amsfonts}
\usepackage{mathtools}
\usepackage{graphicx}
\usepackage{epstopdf}
\usepackage{bm} 
\usepackage{mathrsfs}
\newcommand{\be}{\begin{equation}}
\newcommand{\ee}{\end{equation}}
\newcommand{\bea}{\begin{eqnarray}}
\newcommand{\eea}{\end{eqnarray}}
\newcommand{\ket}{\rangle}
\newcommand{\bra}{\langle}
\newcommand{\mc}{\mathcal}
\providecommand{\mb}{\mathbf}

\begin{document}
\begin{frontmatter}
\title{Chiral Y junction of quantum spin chains}

\author{F. Buccheri$^1$, R. Egger$^1$, R. G. Pereira$^{1,2,3}$ and F. B. Ramos$^2$}
\address{$^1$ Institut f\"ur Theoretische Physik, Heinrich-Heine-Universit\"at, D-40225  D\"usseldorf, Germany}
\address{$^2$ International Institute of Physics, Universidade Federal do Rio Grande do Norte, Campus Universitario, Lagoa Nova, Natal-RN 59078-970, Brazil}
\address{$^3$ Departamento de F\'isica Te\'orica e Experimental, Universidade Federal do Rio Grande do Norte, Natal-RN 59078-970, Brazil}

\begin{abstract}
We study a Y junction of spin-$1/2$ Heisenberg  chains with an interaction that breaks both time-reversal and chain exchange symmetries,
but not their product nor SU(2) symmetry.
The boundary phase diagram features a stable disconnected fixed point at weak coupling and a stable three-channel Kondo fixed point at strong coupling,  separated by an unstable chiral fixed point at intermediate coupling.
Using non-abelian bosonization and boundary conformal field theory, together with density matrix renormalization group  and quantum Monte Carlo simulations, we characterize the signatures of these low-energy fixed points. In particular, we address the boundary entropy, the spin conductance, and the temperature dependence  of the scalar spin chirality  and the   magnetic susceptibility at the boundary.
\end{abstract}


 
\end{frontmatter}

\section{Introduction}\label{sec:Intro}

The field of spintronics has received continued attention
in the last several years \cite{Prinz1998,Wolf2001,awschalom2002semiconductor}.
This is partly due to the discovery of the spin Seebeck effect  and the spin Hall effect,
which allow for the conversion between spin and  charge currents \cite{Uchida2008,Saitoh2006,Maekawa}.
Importantly, the generation and detection of magnonic currents via the spin Hall effect has been recently 
demonstrated 
 using a YIG/Pt heterostructure \cite{Cornelissen2015},
in which the magnons of the ferromagnetic insulator are responsible for spin transport
between the terminals. 
Analogous setups have recently been   realized with antiferromagnetic materials  \cite{Lebrun2018} and
applications to magnon-based computation have   been proposed 
in a multi-terminal platform \cite{Ganzhorn2016}. 
 The essential feature of these experiments  is the presence of
magnetization-carrying excitations with a relaxation length larger than the 
separation between the terminals. In this respect, the antiferromagnetic
  Heisenberg chain represents an idealized case of a one-dimensional (1D) 
  system in which spinful excitations can propagate ballistically   in the low-temperature limit \cite{Meier2003,Lange2018}.
  Moreover, it was shown to describe accurately the behavior of effectively 1D crystals such as CsCoCl$_3$ \cite{Nagler1982,Nagler1983}, CsCoBr$_3$ \cite{Yoshizawa1981,Goff1995}, KCuF$_3$ \cite{Tennant1995}, and 
  Sr$_2$CuO$_3$ \cite{Motoyama1996,Takigawa1996}.
The spectrum of the model has been known for a long time by exact methods \cite{Bethe1931,Takahashi},
and its low-energy properties are essentially captured by field theory \cite{Gogolin2004}. 

A key element of any quantum or classical circuit is a Y
junction, a topic of established interest in condensed matter and
statistical physics \cite{Lal2002,Chen2002,Chamon2003,Barnabe2005,Oshikawa2006,Hou2008,Agarwal2009,Giuliano2009,Aristov2011,Tsvelik2013}.
In this work, we study a   Y junction of spin-$\frac{1}{2}$ chains with a boundary three-spin interaction analogous to the chiral 
spin liquid order parameter proposed in \cite{Wen1989,Baskaran1989}.
This coupling preserves $\mathbb{Z}_3$ cyclic permutation of the chains  as well
as the spin SU(2) symmetry. On the other hand, it  breaks time-reversal
and leg-exchange (reflection) symmetries, while preserving the symmetry under
the combined action of the two. The main question is how this chiral interaction affects the spin transport properties of the Y junction. More specifically, our goal is to identify an interaction regime in which a Y junction made of long chains can redirect spin currents in a  clockwise or counter-clockwise manner, in analogy with quantum circulators that have been realized in photonic and superconducting  circuits \cite{Scheucher2016,Chapman2017,Mahoney2017}. 

The low-energy, long-wavelength physics of the Y junction is governed by conformally invariant boundary conditions \cite{Cardy1986,Cardy1989} connecting the collective spin modes in different chains.  In our earlier contribution \cite{Buccheri2018}, we showed that the Y junction  model discussed below exhibits a remarkable boundary phase diagram as a function of the boundary  interaction strength.
It features two stable  fixed points, corresponding to disconnected chains and to a  three-channel-Kondo fixed point.
The two fixed points are
separated by an unstable chiral fixed point. The latter  realizes an ideal  quantum spin circulator
but requires parameter fine tuning \cite{Buccheri2018}. The experimental realization of such a spin circulator
 would be of practical interest in the field of antiferromagnetic spintronics, aiming
 at the realization of circuits using    Mott insulators \cite{Wadley2016,Jungwirth,Baltz2016}.

Controlled and tunable setups for the simulation of quantum chains are also
provided by cold-atom experiments, which have proven  useful tools for 
studying  Heisenberg chains and multi-spin exchange interactions \cite{Boll2016,Dai2017}.
An interesting perspective on the realization of a  junction of quantum chains
in these platforms is the use of `synthetic dimensions' 
encoded by internal atomic states. This technique  has been
 applied  successfully for creating spin ladders with a synthetic magnetic flux \cite{Celi2014,Livi2016}.

In this paper, we apply non-abelian bosonization and boundary conformal field theory (BCFT) to characterize physical observables 
of Y junctions  of spin chains 
 in the vicinity of  the above fixed points. Going significantly beyond the results reported in our earlier work \cite{Buccheri2018}, here we present a detailed derivation of the boundary entropy and of the spin conductance matrix for each of the three fixed points. Moreover, we present  numerical  results for the spin conductance obtained by density matrix renormalization group (DMRG) methods, and for finite-temperature properties from  quantum Monte Carlo (QMC) simulations.  Put together, these numerical results strongly support  our analytical predictions, in particular the existence of a  chiral fixed point at intermediate coupling. Despite its instability, the chiral fixed point governs the physics of  Y junctions at finite length scales and/or at finite temperatures over a wide parameter regime. 
 
The remainder of this  article is organized as follows. After introducing our model for the Y junction in section
\ref{sec:The-model}, we discuss the bosonization approach and the boundary phase diagram in section \ref{sec:bosonization}. The BCFT formalism is presented in
section \ref{sec:Fusion}, 
with particular attention to the operator content at the various fixed points.
Observables characterizing the different fixed points  are discussed in section \ref{sec:FixedPoints}. We then present our numerical DMRG and QMC
results   in section \ref{sec:Numerical-checks}. Finally, we present our conclusions
in section \ref{sec:Conclusions}.  Several technical details have been delegated to the Appendix.

\section{Model}\label{sec:The-model}

We start with  three decoupled isotropic  spin chains, each labeled by a leg index $\alpha=1,2,3$. The Hamiltonian for a single chain   includes the usual antiferromagnetic nearest-neighbor exchange interaction $J_{1}>0$   supplemented by a next-nearest-neighbor interaction $J_{2}$,
\begin{equation}
H_{\alpha}=J_{1}\sum_{j=1}^{\mc L-1}{\bf S}_{j,\alpha}\cdot{\bf S}_{j+1,\alpha}+J_{2}\sum_{j=1}^{\mc L-2}{\bf S}_{j,\alpha}\cdot{\bf S}_{j+2,\alpha}\label{J1J2}.
\end{equation}
Here $\mathbf S_{j,\alpha}$ is a spin-$\frac{1}{2}$ operator that acts in the Hilbert space of  site $j$ in chain $\alpha$, and $\mc L$ is the length of the chains with open boundary conditions. We shall be mainly interested in the limit of semi-infinite chains, $\mc L\to\infty$, but see sections \ref{sec:Fusion} and \ref{subsec:DMRG}. In the presence of a $\mathbb{Z}_3$ cyclic chain permutation symmetry, which takes $\alpha\mapsto \alpha+1 $ (mod 3), the interaction parameters $J_1$ and $J_2$  are constrained to be the same for the three chains. In addition, $H_\alpha$ is invariant under time reversal, acting on spin operators as\be
\mc T:\mathbf S\mapsto -\mb S,
\ee
and under the $\mathbb Z_2$ leg exchange  (say, of chains 1 and 2) \be
\mc P:\alpha \mapsto -\alpha \textrm{ (mod 3)}.
\ee

\begin{figure}
\begin{centering}
\includegraphics[width=0.35\textwidth]{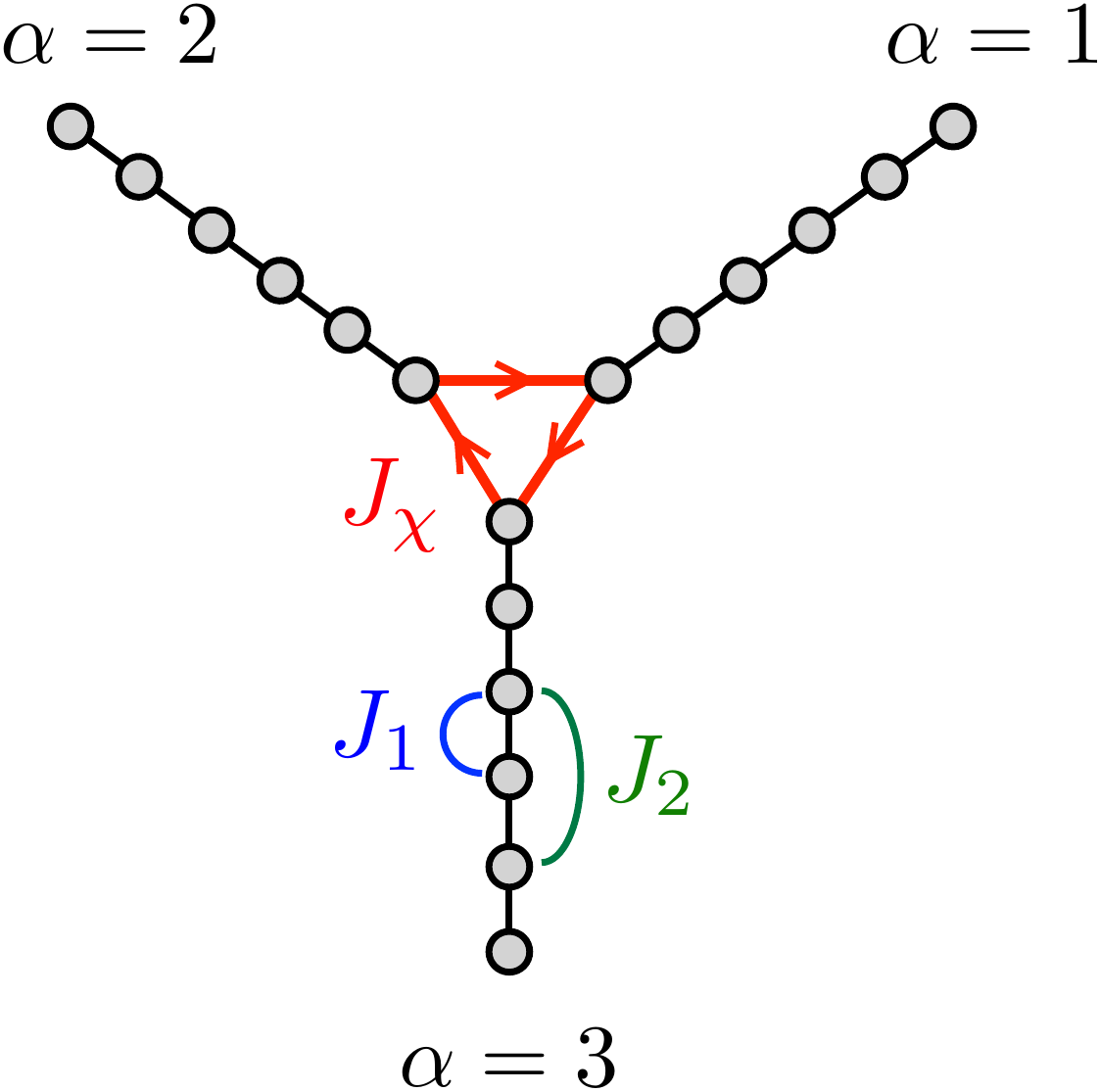}
\caption{Y junction of spin chains with bulk  exchange  
couplings $J_1$ and $J_2$ and boundary three-spin interaction $J_\chi$. 
The arrows indicate the chirality favored by $J_\chi>0$. \label{fig1}}
\par\end{centering}
\end{figure}

The ground-state phase diagram of the $J_1$-$J_2$ chain in  (\ref{J1J2}) is well known \cite{Majumdar1969,Haldane1982}. There is a critical value $J_{2}^{c}\approx 0.2412J_{1}$ \cite{Okamoto1992,Eggert1992,Laflorencie2008}   separating a critical phase for $0\leq J_2\leq J_2^c$ from a gapped dimerized phase for $J_2>J_2^c$. We focus on the critical phase and add to the Hamiltonian  the most general boundary interactions, coupling only  spins at position $j=1$ in each chain. These interactions are allowed to break   $\mc T$ and  $\mathcal{P}$
symmetries but are required to preserve  SU(2), $\mc P\mc T$  and  
  $\mathbb{Z}_{3}$   symmetries. The full Hamiltonian
takes the form 
\begin{equation}
H=\sum_{\alpha=1}^{3}H_{\alpha}+H_{B},\label{Htotal}
\end{equation}
 with the boundary interaction term
 \be 
 H_B=J'\sum_{\alpha=1}^3\mathbf{S}_{1,\alpha}\cdot\mathbf{S}_{1,\alpha+1}+J_\chi \hat C_1.\label{HB}
\ee
Here we define the scalar spin chirality operator (SSCO)
as \cite{Wen1989}
\be
\hat{C}_{j}=\mathbf{S}_{j,1}\cdot(\mathbf{S}_{j,2}\times\mathbf{S}_{j,3}).\label{SSCO}
\ee
The first term in (\ref{HB}) is a simple exchange interaction between the end spins. This term preserves both $\mc P$ and $\mc T$ symmetries but by itself cannot lead to chiral spin transport through the junction. The second term involves  the SSCO for the end spins. We note that expectation values of SSCOs have been proposed as order
parameters \cite{Wen1989,Baskaran1989} for characterizing  chiral
spin liquid phases \cite{Kalmeyer1987,Kalmeyer1989}. The $J_\chi$ term in (\ref{HB}) can be derived
starting from  the Hubbard model in the limit of strong
on-site repulsion and treating virtual hopping processes in the presence of a magnetic flux  \cite{Sen1995}. Alternatively, it  appears in effective Floquet spin models for a Mott insulator driven by circularly polarized light \cite{Kitamura2017,Claassen2017}.
 Since we are mostly interested in studying the possibility of chiral spin transport, hereafter we set $J'=0$ and retain only the  boundary parameter $J_\chi$, see figure \ref{fig1}. 

\section{The boundary phase diagram \label{sec:bosonization}}

In the critical phase $0\leq J_2\leq J_2^c$,  the low-energy effective field theory
describing the continuum limit of Hamiltonian (\ref{J1J2}) is the SU(2)$_{1}$
Wess-Zumino-Witten (WZW) model 
 with an open boundary at the origin \cite{Gogolin2004},
\begin{equation}
H_{\alpha}=\int_{0}^{\infty}dx\left\lbrace \frac{2\pi v}{3}\left[\mathbf{J}_{\alpha,L}^{2}\left(x\right)+\mathbf{J}_{\alpha,R}^{2}\left(x\right)\right]-2\pi v\gamma\mathbf{J}_{\alpha,L}(x)\cdot\mathbf{J}_{\alpha,R}
\left(x\right)\right\rbrace \label{continuumlimitchain}.
\end{equation}
Here, $v$ is the spin velocity and $\gamma\geq 0$ is the dimensionless coupling constant of the  marginally irrelevant  operator.  The critical point $J_2=J_2^c$ corresponds to $\gamma=0$  in the effective
Hamiltonian (\ref{continuumlimitchain}). At this point, the model becomes equivalent to free bosons
(up to strictly irrelevant perturbations) and  logarithmic corrections to correlations functions vanish \cite{Eggert1992}. In order to simplify our analysis
and focus on the essential physics of the Y  junction, we will henceforth consider
the bulk of the spin chains to be tuned to this critical point.  The chiral spin current operators $J_{\alpha, L}^{a}(z)$   and $J_{\alpha, R}^{a}(\bar z)$ (with $a=1,2,3$) represent  collective spin modes moving in the direction of decreasing or increasing $x$, respectively. They are functions of the complex
coordinates $z=v\tau-ix$ or $\bar{z}=v\tau+ix$,  where $\tau$ denotes imaginary time.
These chiral spin currents are not be independent; instead, their relation is fixed by the boundary conditions, which have not yet been specified in (\ref{continuumlimitchain}).  
 With our normalization choice, their operator product
expansion (OPE) satisfies the SU(2)$_{k}$ Kac-Moody algebra
\cite{di1997conformal},
\begin{equation}
J_{\alpha,L}^{a}\left(z\right)J_{\alpha,L}^{b}\left(w\right)
\sim \frac{1}{8\pi^2}\frac{k\delta_{ab}}{\left(z-w\right)^{2}}+\frac{1}{2\pi}\frac{i\varepsilon_{abc}}{z-w}J_{\alpha,L}^{c}\left(w\right),\label{KMope}
\end{equation}
with $k=1$ and the Levi-Civita tensor $\varepsilon$.
 The boundary conditions will be determined as a function of $J_\chi$ in (\ref{HB}) according to a `delayed evaluation of boundary conditions' \cite{Oshikawa2006}. The boundary term $H_B$ acting at $x=0$ will determine the boundary conditions, which in turn fix the dependence of the right from the left currents and the full set of OPEs.

The effective field theory can be formulated in terms of just one  (left- or right-moving) boson field for each chain $\alpha$.
These chiral boson fields, $\varphi_{\alpha,\mu}(x)$ with $\mu=L/R=+/-$, obey the commutation relations
\begin{equation}
\left[\varphi_{\alpha,\mu}\left(x\right),\partial_{x'}\varphi_{\alpha',\mu' }\left(x'\right)\right]=i\mu\delta_{\alpha\alpha'}\delta_{\mu\mu'
}\delta\left(x-x'\right).
\label{chiralvarphicomm}
\end{equation}
One can write the currents and the spin operators, as well as the boundary SSCO, in terms of these fields (see \ref{app:WZW}). Without imposing boundary conditions yet, we write the boundary interaction as
\begin{eqnarray}\label{cboson}
H_B&\simeq& \frac{J_\chi}{\pi^2}\sum_\alpha \left(\frac{1}{2\sqrt{\pi}}
\left(\partial_x \varphi_{\alpha,L} -\partial_x \varphi_{\alpha,R}\right)+ \mathcal{A}
\sin\left[\sqrt{\pi}\left(\varphi_{\alpha,L}-\varphi_{\alpha,R}\right)\right]\right)
\nonumber\\
&&
\times \sin\left[\sqrt{\pi}\left( \varphi_{\alpha+1,L}+ \varphi_{\alpha+1,R}-\varphi_{\alpha-1,L}- \varphi_{\alpha-1,R}\right)\right] 
\\
   &&
\times \left(\pi\mathcal{A}+ \cos\left[\sqrt{\pi}\left(\varphi_{\alpha+1,L}-\varphi_{\alpha+1,R}\right)\right]\right)
\nonumber\\
&&
\times \left(\pi\mathcal{A}+ \cos\left[\sqrt{\pi}\left(\varphi_{\alpha-1,L}-\varphi_{\alpha-1,R}\right)\right]\right)\;,
\nonumber
\end{eqnarray}
where the real positive number $\mathcal{A}$ is a nonuniversal constant of order unity.

\begin{figure}
\begin{centering}
\includegraphics[width=0.8\textwidth]{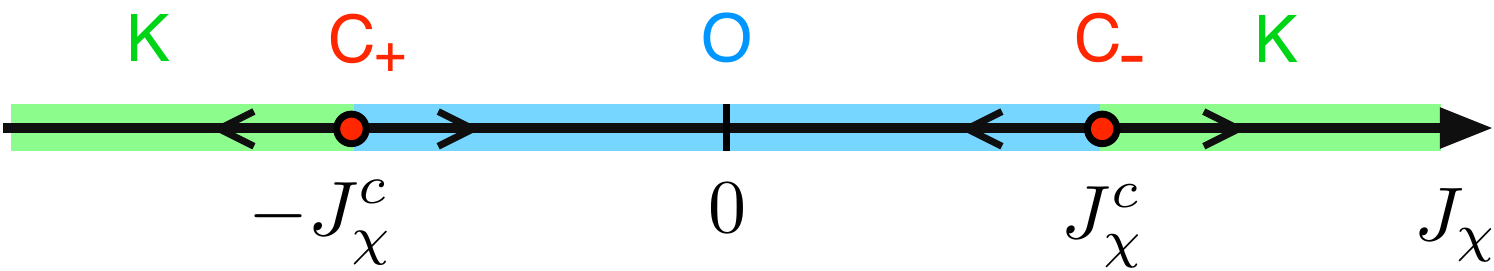}
\caption{Schematic boundary phase diagram. 
The stable fixed points have emergent time-reversal symmetry and correspond to open 
(O) boundary conditions at weak coupling and a three-channel Kondo fixed point (K) 
at strong coupling. At the critical  values $J_\chi=\pm J_\chi^c$, the Y junction 
is described by  unstable chiral fixed points with either clockwise or 
counter-clockwise circulation of spin currents (C$_-$ or C$_+$, respectively).  
\label{fig2}}
\par\end{centering}
\end{figure}


 In \cite{Buccheri2018}, as  $J_\chi$ is varied, we have argued for the phase diagram shown in figure \ref{fig2}. First, we note that the fixed point of disconnected chains with open boundary conditions (O fixed point). The open boundary condition 
 \be
 \varphi_{\alpha,R}(x)=\varphi_{\alpha,L}(-x)+\mc C,\label{bcbosons}
 \ee
 where the constant $\mc C$ is constrained to $\mc C=0$ or $\mc C=\pm\sqrt\pi$ by virtue of SU(2) invariance, is stable for $|J_\chi|$ below a critical value $J_\chi^c$. In fact, after imposing (\ref{bcbosons}) we obtain the boundary SSCO in the form
\be
\hat C_1\propto   \sum_\alpha \partial_x\varphi_{\alpha,L}(0)\sin[\sqrt{4\pi}\varphi_{\alpha+1,L}(0)-\sqrt{4\pi}\varphi_{\alpha-1,L}(0)].
\ee
This is a dimension-three boundary operator that represents a highly irrelevant perturbation near the O point. In non-abelian bosonization, we can write the boundary SSCO as the triple product of the chiral spin currents, 
\be
\hat C_1\propto \sum_\alpha \mathbf J_{\alpha,L}(0)\cdot [\mb J_{\alpha+1,L}(0)\times \mb J_{\alpha-1,L}(0)].
\ee 
This   SU(2)-symmetric form can be obtained directly from equation (\ref{SSCO}) by using that, in the continuum limit with open boundary conditions, $\mathbf S_{\alpha,j=1}\propto \mb J_{\alpha,L}(0)$ \cite{Eggert1992}.   Beyond first order in $J_\chi$, we must take into account all boundary operators allowed by symmetry, as they can be generated by the renormalization group (RG) flow.  The leading boundary operators that perturb the O fixed point are given by
\bea
H_B^{(O)}&=&\lambda_1^{(O)}\sum_{\alpha}\mb J_{\alpha,L}(0)\cdot \mb J_{\alpha+1,L}(0)+\lambda_2^{(O)}\sum_{\alpha}[\mb J_{\alpha,L}(0)]^2\nonumber\\
&&+\lambda_3^{(O)}\sum_\alpha \mathbf J_{\alpha,L}(0)\cdot [\mb J_{\alpha+1,L}(0)\times \mb J_{\alpha-1,L}(0)]+\cdots,
\label{HBO}
\eea
where $\cdots$ refers to more irrelevant operators. While $\lambda_3^{(O)}\sim J_\chi$, the leading irrelevant boundary operator that couples the chains is the time-reversal-invariant boundary exchange coupling $\lambda_1^{(O)}$, a dimension-two operator generated by the RG to second order in $J_\chi$ (or first order in $J'$ in the more general model). 

It is important to underline that  the argument for the stability of the O fixed point is only perturbative in the coupling to the boundary SSCO. For  $|J_\chi|\sim J_1$, we expect to find a different boundary condition whereby the chiral bosonic modes $\varphi_{\alpha,R}$ are connected with the modes $\varphi_{\alpha\pm1,L}$ in the other  chains. Following  \cite{Chamon2003,Oshikawa2006,Bellazzini2007}, we consider  conformally invariant boundary conditions of the form
\begin{equation}
\varphi_{\alpha,R}\left(x\right)=\sum_{\beta=1}^{3}M_{\alpha\beta}\varphi_{\beta,L}\left(-x\right)+\mc C,\label{linearBC}
\end{equation}
where $M$ is a $3\times 3$ matrix and $\mc C$ is a constant. At the O fixed point, $M$ is simply the identity.   More
generally, the condition of preserving the algebra in equation (\ref{chiralvarphicomm})
implies that $M$ must be orthogonal.  The orthogonality of $M$ also implies that the scaling
dimension of the chiral currents is not modified by the boundary condition. Imposing SU(2) symmetry for the Y junction  leads to further constraints. First, we must have either $\mc C=0$ or $\mc C=  \sqrt{\pi}$.  Second, the leading terms in the OPE of $J^+_{\alpha,R}(x)J^-_{\alpha\pm1,L}(x')$ must have the same scaling dimension as $J^z_{\alpha,R}(x)J^z_{\alpha\pm1,L}(x')$, namely dimension two. One can conclude, using the bosonization formulas of the spin operators (\ref{J2phi}), that the only options that are compatible with the above conditions are as follows. (i) $M_{\alpha\beta}=\delta_{\alpha,\beta}$, which corresponds to open boundary conditions, (ii) $M_{\alpha\beta}=\delta_{\alpha,\beta-1}$, a cyclic permutation in one direction, or (iii) $M_{\alpha\beta}=\delta_{\alpha,\beta+1}$, a cyclic permutation in the other direction. The latter two options correspond to  chiral boundary conditions,
\bea
\textrm{C}_+:\,\varphi_{\alpha+1,R}(x)=\varphi_{\alpha,L}(-x)+\mc C,\label{chiralbc1}\\
\textrm{C}_-:\,\varphi_{\alpha-1,R}(x)=\varphi_{\alpha,L}(-x)+\mc C.\label{chiralbc2}
\eea
In terms of chiral currents, we have
\bea
\textrm{C}_{\pm}: \,\mb J_{\alpha\pm1,R}(x)=\mb J_{\alpha,L}(-x).\label{bcchiralJ}
\eea
At the $\textrm{C}_\pm$ fixed points, the leading term in the OPE of the boundary SSCO is proportional to the identity. This is consistent with the fact that the boundary interaction in  (\ref{cboson}) is responsible for pinning the bosonic fields through chiral boundary conditions, similarly to other applications of the method of delayed evaluation of boundary conditions \cite{Oshikawa2006,Bellazzini2009,Mardanya2015}.  As a result,  we find  a nonzero expectation value,
\be
\textrm{C}_\pm:\,\langle \hat C_{j}\rangle=\pm(-1)^j \frac{\mc A^3}2\cos(\sqrt\pi \mc C),
\ee
for sites $j$ near the boundary, i.e., where we can approximate $\varphi_{\alpha,L}(x)\approx \varphi_{\alpha,L}(-x)$ for $x\to0$ in the slowly varying fields. The choice of $\mc C=0$ or $\mc C=  \sqrt \pi$ thus determines whether the scalar spin chirality is positive on even or odd sites near the boundary. For the boundary interaction acting at site $j=1$, we obtain the expectation value 
\be
\textrm{C}_\pm:\,\langle H_B\rangle=\mp \frac{J_\chi\mc A^3}2\cos(\sqrt\pi \mc C). \label{energy}
\ee
Which chiral fixed point is then realized will depend on the sign of $J_\chi$. For instance, $\textrm{C}_-$ with $\mc C=  \sqrt{\pi}$ corresponds to the case where the incoming spin current from chain $\alpha$ is channeled  into chain $\alpha-1$, and $\langle \hat C_1\rangle<0$. We expect this fixed point to be selected for a boundary interaction  (\ref{HB}) with $J_\chi>0$, since this choice lowers the ground state according to (\ref{energy}).  The spin current then circulates clockwise in figure \ref{fig1}, as  expected for a local negative-chirality state with  ordering $3\to2\to1\to3$. 
This hypothesis can be checked by numerically calculating the spin conductance and the three-spin correlations, see section \ref{sec:Numerical-checks}.

We next analyze   the stability of the chiral fixed points in the bosonization approach by considering other terms generated in the OPE of the boundary SSCO besides the identity. Imposing chiral boundary conditions in (\ref{cboson}), we obtain 
\bea\label{CnJJcbc}
\textrm{C}_\pm :\,H_B&& \simeq \mp \frac{J_\chi}2\cos(\sqrt\pi \mc C)\\ \nonumber
&&\times\left\{\mc A^3+\frac{\mc A}{4\pi^2}\sum_{\alpha}\cos[\sqrt{\pi}\varphi_{\alpha,L}(0)-\sqrt{\pi}\varphi_{\alpha+1,L}(0)]\right\}+\cdots,
\eea
where we omit irrelevant operators. The cosine term  in  (\ref{CnJJcbc})
has scaling dimension $1/2$ and thus corresponds to a relevant boundary operator. We can  interpret it as being due to the backscattering of spin currents, in analogy to the relevant
backscattering operator in a Y junction of quantum wires with Luttinger parameter $K=1/2$ (as appropriate for  Heisenberg chains) \cite{Oshikawa2006}. In essence, the relevant boundary operator stems from the fact that for chiral boundary conditions, the boundary spin operator  (\ref{nabSpin}) does not reduce to the chiral currents $\mb J_{\alpha,L}(0)$ but also contains a contribution proportional to the dimension-$1/2$ matrix field.   Indeed, the most relevant perturbation to the chiral fixed point is given by
\be
H_B^{(C)}=\lambda_1^{(C)}\sum_\alpha\textrm{Tr}[\tilde g_\alpha(0)]+\cdots,\label{HBchiral}
\ee
where  $\tilde g_\alpha(x)$ are matrix fields  obtained from the original $g_\alpha(x)$ by replacing $\varphi_{\alpha,R}(x)\mapsto \varphi_{\alpha\pm1,L}(-x)$, cf.~equation (\ref{dimer}).  If we choose the boundary conditions such that the prefactor in (\ref{CnJJcbc}) is negative, we will have  $\lambda_1^{(C)}<0$  with $|\lambda_1^{(C)}|\sim |J_\chi|$ at weak coupling.  As a consequence, the chiral fixed points cannot appear at weak coupling as they are destabilized by a relevant perturbation. 

However, the existence of only one relevant operator in  (\ref{HBchiral}) suggests that  chiral fixed points can be realized  as critical points at intermediate coupling, reached by tuning a single parameter in the lattice model. Let us suppose that there exists a critical value   $J_\chi^c>0$, where the relevant coupling vanishes and  $\lambda_1^{(C)}\approx -b(|J_\chi|-J_\chi^c)$ for $|J_\chi|\approx J_\chi^c$.   We must have $b>0$ so that $\lambda_1^{(C)}<0$ for  $|J_\chi|<J_\chi^c$ as verified at weak coupling. What happens when $\lambda_{1}^{(C)}$ changes sign across the putative critical point? Under the RG,  the effective coupling   $\lambda_1^{(C)}<0$ flows monotonically  to strong coupling $ \lambda_1^{(C)}\to -\infty$ at low energies. When this happens,
the boson fields are pinned to the minima of the cosine potentials in  (\ref{CnJJcbc}),
\be
\varphi_{\alpha,L}(0)-\varphi_{\alpha+1,L}(0)=2n_\alpha\sqrt\pi,\qquad n_\alpha\in \mathbb Z. \label{Backscatteringpinning}
\ee
On the other hand,  for $\lambda^{(C)}_{1}>0$ the effective coupling flows to $ \lambda_1^{(C)}\to +\infty$ under the RG, and the cosine potentials pin the boson fields to 
\be
\varphi_{\alpha,L}(0)-\varphi_{\alpha+1,L}(0)=(2n_\alpha+1)\sqrt\pi,\qquad n_\alpha\in \mathbb Z. \label{BackscatteringpinningNegative}
\ee
We can understand the difference between  equations (\ref{Backscatteringpinning})
and (\ref{BackscatteringpinningNegative}) by noting that when we impose chiral boundary conditions, the smooth part of the spin operator is represented by 
\bea
\textrm{C}_\pm:\,S_{j,\alpha}^z&\sim&J^z_{\alpha,L }(x)+J^z_{\alpha\mp1,L }(-x) \nonumber \\
&=& \frac{1}{\sqrt{4\pi}}\left[\partial_x\varphi_{\alpha,L}(x) +\partial_x\varphi_{\alpha\mp1,L}(-x) \right]. 
\eea
The magnetization measured out to large distances from the boundary is
\bea
  \sum_{j=1}^\infty S_{j,\alpha}^z&\sim &\frac{1}{\sqrt{4\pi}}\left[ \int_0^\infty dx\,\partial_x\varphi_{\alpha,L}(x) -\int_0^\infty dx'\,  \partial_{x'}\varphi_{\alpha\mp1,L}(x')\right]\nonumber\\
&=&\frac{1}{\sqrt{4\pi}} \left[ \varphi_{\alpha\mp1,L}(0)-\varphi_{\alpha,L}(0) \right].
\eea
The pinning conditions   (\ref{Backscatteringpinning})
and (\ref{BackscatteringpinningNegative}) thus correspond to
\begin{equation}
 \sum_\alpha \sum_{j=1}^\infty S_{j,\alpha}^{z}  =\left\{\begin{array}{cc}
n, & \quad\lambda^{(C)}_{1}\to-\infty\\
n+\frac{1}{2}, & \quad\lambda_{1}^{(C)}\to+\infty
\end{array}\right.,
\end{equation}
where $n\in\mathbb{Z}$.  This suggests that changing the sign of $\lambda_{1}^{(C)}$ from negative to positive  brings an effective spin $1/2$ from infinity
to the boundary. A similar effect occurs with  the backscattering potential $\lambda\cos [\sqrt{4\pi K}\phi (0 )]$ for the Kane-Fisher problem of a single impurity in a Luttinger liquid  \cite{Kane1992}. For repulsive interactions, corresponding to Luttinger parameter $K<1$, the backscattering operator is relevant for both signs of
the coupling $\lambda$. The critical point $\lambda=0$ corresponds to a resonant tunneling condition, where the number of electrons on the scattering center fluctuates \cite{Kane1992}. Changing the sign of the coupling   switches between a predominantly repulsive and a predominantly attractive scattering potential. In the
latter case, a bound state is formed for arbitrarily weak scattering
amplitude, and  exactly one extra   unit of charge is  bound near the origin. By analogy, we expect that in our   SU(2)-symmetric spin chain  problem the sign change   of $\lambda_1^{(C)}$ entails the formation of  an effective spin-$1/2$ degree of freedom at the boundary  for $|J_\chi|>J_\chi^c$. 

An analysis of our Y junction model in the strong coupling limit corroborates the above picture.  For $|J_{\chi}|\gg J_1$, one
may, in a first approximation, neglect the coupling of the three end
spins to the rest of the chain and diagonalize the reduced Hamiltonian in (\ref{HB}), where we now include $J'\neq 0$. 
The spectrum of this three-spin system can be labeled by the quantum numbers of 
total spin, $s$, magnetization, $s^z$, and the scalar spin chirality, $c$.
The spectrum is composed of a non-chiral
($c=0$) spin-$3/2$ quadruplet at energy $3J'/4$ and two
spin-$1/2$ doublets with opposite chiralities ($c=\pm1$) and energies 
\begin{equation}\label{SCenergies}
E_c\left(J_{\chi},J'\right) =-\frac{3J'}{4}+c J_{\chi}\frac{\sqrt{3}}{4} .
\end{equation}
As long as $J'>-\frac{|J_\chi|}{2\sqrt3}$, the ground state of $H_B$ is twofold degenerate. For $J_\chi>0$,  the ground state  is  the doublet with
negative chirality, given by the states \cite{Wen1989}
\begin{eqnarray}
 \left|s=\frac{1}{2},s^z=\frac{1}{2},c=-1\right\rangle &=&\frac{i}{\sqrt{3}}\left(\left|\downarrow\uparrow\uparrow\right\rangle+\omega\left|\uparrow\uparrow\downarrow\right\rangle +\omega^2\left|\uparrow\downarrow\uparrow\right\rangle  \right),\nonumber\\
 \left|s=\frac{1}{2},s^z=-\frac{1}{2},c=-1\right\rangle &=&-\frac{i}{\sqrt{3}}\left(\left|\uparrow\downarrow\downarrow\right\rangle+\omega^2\left|\downarrow\uparrow\downarrow\right\rangle+\omega\left|\downarrow\downarrow\uparrow\right\rangle \label{chiralityeigenstates}   \right),
\end{eqnarray}
where $\omega=e^{i2\pi/3}$ and $\omega^2=\omega^*=\omega^{-1}$. Both states are conjugated by $\mc{PT}$  with $(\mc{PT})^2=-1$,
and therefore their degeneracy is protected by $\mc{PT}$ symmetry. 
For $J_\chi<0$, the ground state is given by the positive-chirality states obtained from  (\ref{chiralityeigenstates})  by applying either $\mc P$ or $\mc T$. 

We can  now define an effective spin-$1/2$ operator, $\mathbf{S}_{0}$, 
which acts in the Hilbert subspace spanned by the two states  (\ref{chiralityeigenstates}). For $J_\chi\gg J_1$, we project the full Hilbert space of the boundary spins 
  onto this low-energy subspace.  The projection of the total Hamiltonian leads to \begin{equation}
H_{\textrm{\tiny eff}}=\sum_{\alpha}\tilde H_\alpha+J_{K}\mathbf{S}_{0}\cdot\sum_{\alpha}\left(\mathbf{S}_{2,\alpha}+\frac{J_2}{J_1}\mb S_{3,\alpha}\right),
\end{equation}
where $J_K=J_1/3$ and 
\be
\tilde H_\alpha=\sum_{j\geq2} \left(J_1\mb S_{j,\alpha}\cdot \mb S_{j+1,\alpha}+J_2  \mb S_{j,\alpha}\cdot \mb S_{j+2,\alpha}\right)
\ee 
is the new Hamiltonian for each chain where the site $j=1$ has been removed to form the central spin $\mb S_0$. 
Remarkably, this projected Hamiltonian has an emergent time-reversal symmetry since it only contains two-spin (exchange) interactions. However, time-reversal symmetry breaking perturbations appear in the effective Hamiltonian to leading order in $J_1/J_\chi$ once we take into account virtual transitions to the excited states of $H_B$ \cite{Buccheri2018}. 

We note that the coupling $J_K$ between the central spin $\mb S_0$   and the new boundary  spins   of the   chains is suppressed by a factor 1/3 in comparison with the bulk exchange coupling. Thus, we can approach the Y junction in the strong coupling limit as the problem of an impurity spin weakly  coupled to three chains with open boundary conditions at $j=2$. In contrast with the O fixed point, this new starting point is unstable because $J_K>0$ is equivalent to an antiferromagnetic Kondo coupling which is marginally relevant. Taking the continuum limit of the projected Hamiltonian, we obtain (for $J_2=J_2^c$)
\be
H_{\textrm{\tiny eff}}\approx\sum_\alpha\int_{-\infty}^{\infty}dx\,  \frac{2\pi v}{3} \mb J_{\alpha,L}^2(x) -2\pi v \lambda_K \mb S_0\cdot \mb J_{\alpha,L}(0),
\ee
where $\lambda_K\propto J_{K}/(2\pi v)$ is the dimensionless coupling constant. 
The perturbative RG equation takes the form \cite{Laflorencie2008,Barzykin1998}
\begin{equation}
\frac{d\lambda_{K}}{d\,l}=\lambda_{K}^{2}-\frac{3}{2}\lambda_{K}^{3}+\cdots,\label{strongcouplingrg}
\end{equation}
where $l=\ln(\Lambda_0/\Lambda)$ with  $\Lambda$ denoting the running high-energy cutoff. 
This equation coincides with the RG equation for the three-channel Kondo effect with itinerant  electrons \cite{Nozieresph,hewson1997kondo,Gogolin2004}.  Even though our model lacks the charge degrees of freedom of the original  Kondo problem, it is known that the single-channel and two-channel Kondo effects can be realized with spin chains \cite{Eggert1992, Laflorencie2008,Alkurtass2016}. Here we extend this analogy to the case of three spin chains and identify a low-energy three-channel Kondo (K) fixed point  in the regime  $|J_\chi|\gg J_1$. It is also interesting to note that the scenario is connected to the results of \cite{Tsvelik2013b}, with the important difference that here the impurity is connected to the end of the spin chains. In order to provide a unified description of all fixed points of the Y junction, we now turn to the BCFT approach. 

\section{Non-abelian bosonization and boundary conformal field theory}\label{sec:Fusion}

The key assumption in the BCFT approach is that the stable fixed points  are described by conformally invariant boundary conditions, which can be generated using the  fusion algebra of  the primary fields of the theory  \cite{Cardy1986,Cardy1989,Cardy2004}. This approach has been applied successfully to several quantum impurity problems, e.g., the multi-channel Kondo model \cite{Affleck1991,AffleckLudwig2,Affleck1993b}, junctions of two spin chains \cite{Affleck1993}, or electronic Y junctions \cite{Oshikawa2006,Hou2008}. In the following, we apply  
BCFT to describe the operator content at the fixed points of a Y junction of spin chains.

\subsection{Conformal embedding and partition function}

The first step will be to identify a suitable conformal embedding \cite{di1997conformal,Zamolodchikov1986} which can be used to describe the system at low temperatures. Here, we can proceed in analogy to the case of two spin chains \cite{Eggert1992,Affleck1993}. 
The field theory for three decoupled spin chains at low energies is a CFT with total central charge $c=3$,
arising from three copies of a spin SU(2)-symmetric model. The boundary interaction, $H_B$ with $J_\chi\neq 0$, breaks the symmetry to SU(2)$\times\mathbb Z_3$, as only the total spin of the three chains is conserved. The sum of three SU(2)$_1$ chiral currents,
\be
\mc J^a(z)=\sum_{\alpha=1}^3J^a_{\alpha,L}(z),\label{J0current}
\ee
naturally defines the generator of an SU(2)$_3$ Kac-Moody algebra, where $\mc J^a(z)$ obeys equation (\ref{KMope}) at level $k=3$. The SU(2)$_{3}$ WZW field theory 
has central charge $c=9/5$ \cite{di1997conformal,Zamolodchikov1986}.

While the SU(2)$_3$ WZW model describes the total spin degree of freedom, we also need to associate a CFT to 
the `flavor' (or `channel') degree of freedom  carrying the remaining central charge.
 The $\mathbb{Z}_{3}^{(5)}$ CFT is a representative of a family of minimal models which contain
parafermionic fields \cite{Zamolodchikov1985,Fateev1987,Lukyanov1988}.
In addition to the conformal invariance,
it possesses an additional infinite-dimensional symmetry, known as $W_3$ algebra, generated
by a local current. Its central charge is $c=6/5$, so the conformal embedding SU(2)$_3\times \mathbb{Z}_{3}^{(5)}$ 
has indeed $c_{\textrm{\tiny tot}}=3$.  This embedding was used to study a three-leg spin ladder in \cite{Totsuka1996}.  An alternative embedding, which also gives the correct central charge, employs the Ising and tricritical Ising CFTs \cite{Affleck2001}. However, the boundary conditions generated by fusion in these sectors would not be equivalent in general.  We have been able  to reproduce the properties of all three fixed points identified  in section \ref{sec:bosonization} only with the SU(2)$_3\times \mathbb{Z}_{3}^{(5)}$ embedding.

The $\mathbb{Z}_{3}^{(5)}$ theory has $20$ primary fields.  The full list \cite{Fateev1988} can be found in \ref{app:Z3}. The fusion algebra
has been computed from the modular $\mathcal{S}$ matrix and the Verlinde
formula in \cite{frenkel1992}. For convenience, the results are summarized in \ref{app:Z3}. 
Importantly, as a consequence of the underlying symmetry algebra, not all the primary fields appear
at the same time in the partition function, see below. 

\begin{figure}
\begin{centering}
\includegraphics[width=0.65\textwidth]{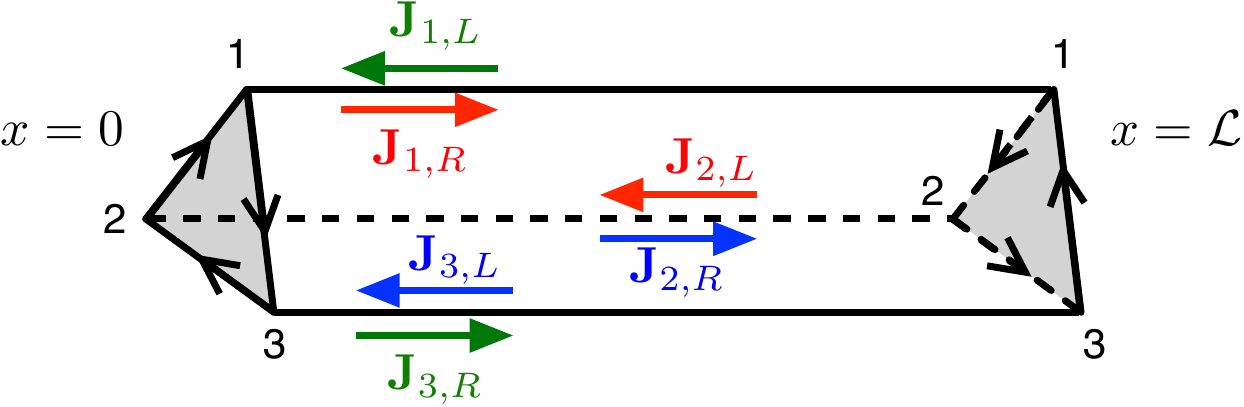}
\caption{Double  Y junction on a finite system with three-spin coupling 
$J_\chi$ at $x=0$ and its mirror image $-J_\chi$ at $x=\mc L$. 
At the chiral fixed point  $J_\chi=J_\chi^c>0$, the chiral currents obey the 
same boundary conditions at both ends:  $\mb J_{\alpha,R}(0)=\mb J_{\alpha+1,L}(0)$ 
and $\mb J_{\alpha,R}(\mc L)=\mb J_{\alpha+1,L}(\mc L)$. 
Modes connected by the boundary conditions are identified by the same color.
\label{fig3}}
\par\end{centering}
\end{figure}

The operator content of two-dimensional (2D) CFTs is organized into infinite-dimensional representations 
generated by a primary field. The characters $\chi$ of a representation
yield a compact way of encoding the operator content of the theory 
\cite{di1997conformal,frenkel1992,Lukyanov1988W}.
In our case, the boundary conditions select a particular subspace of the tensor
product of the spin, SU(2)$_3$, and flavor, $\mathbb{Z}_{3}^{(5)}$, Hilbert spaces. This information is contained in the  partition function, which  has the general form 
\begin{equation}
Z=\sum_{s=0}^{3/2}\sum_{f}n_{sf}\chi_{s}^{\textrm{\tiny SU(2)}_{3}}\left(q\right)\chi_{f}^{\mathbb{Z}_3}\left(q\right),
\label{Zembedding}
\end{equation}
where  $f$ runs over the set of primary fields of  $\mathbb Z_3^{(5)}$,  and the integers $n_{sf}$  determine  
 the operator content for given boundary conditions
\cite{Ludwig1994}. The generally complex parameter $q$ is specialized to
$q=e^{-\frac{\pi\beta}{\mc L}}$, with temperature $T=1/\beta$ and length of a single chain $\mc L$, in order to describe the partition function \cite{Affleck1993}.
Importantly, for sufficiently strong interactions, the boundary degrees of freedom can reorganize the
spectrum in a way encoded by the integers $n_{sf}$.
This is at the base of the \emph{fusion} approach to Kondo
problems and 2D quantum Brownian motion \cite{Affleck1991b,Affleck2001}.
To describe the operator content for semi-infinite chains,  a useful procedure is to map the upper half plane, $(\tau,x)$ with $x\geq0$, onto an infinite strip of width $\mathcal{L}$ \cite{Affleck1993b}. This is accomplished via the transformation
\begin{equation}\label{conformalmap}
z\mapsto w(z)=\frac{\mathcal{L}}{\pi}\ln z.
\end{equation}
In the Y junction, this conformal mapping relates the semi-infinite system to the finite system shown in figure \ref{fig3}, 
where the boundary at $x=\mathcal{L}$ is the mirror image of the boundary at $x=0$ \cite{Rahmani2010,Rahmani2012}. 
This amounts to opposite signs of the boundary coupling, i.e., $J_\chi$ at $x=0$ and $-J_\chi$ at $x=\mathcal{L}$.

We start with the partition function for open boundary conditions, i.e., near the O fixed point. The partition function of decoupled spin chains can
be formulated in terms of the characters of the primary fields of
the SU(2)$_{1}$ WZW CFT \cite{Affleck1993}. Let $\chi_s^{\textrm{\tiny SU(2)}_{1}}(q)$, with $s=0,\frac12$,  denote the character of the spin-$s$ primary fields of SU(2)$_1$. The latter have    dimension $0$  for $s=0$ and $\frac{1}{2}$ for $s=\frac12$. 
 For three open chains with even
($e$) or odd ($o$) values of $\mc L$ (thus, with integer or half-integer values of the total spin), we have the partition
functions 
\begin{equation}
Z_{OO}^{eee}\left(q\right) = \left[\chi_{0}^{\textrm{\tiny SU(2)}_{1}}\left(q\right)\right]^{3},
\qquad\qquad
Z_{OO}^{ooo}\left(q\right) = \left[\chi_{1/2}^{\textrm{\tiny SU(2)}_{1}}\left(q\right)\right]^{3}.
\label{eq:ZOOaff}
\end{equation}
Using the conformal embedding, we can rewrite the partition function in terms of characters of the SU(2)$_{3}$ WZW and  
  $\mathbb{Z}_{3}^{(5)}$ theories. Details are provided in \ref{app:Partitionfunction}, and the final expression is \cite{Totsuka1996}
\begin{eqnarray}
Z_{OO}^{eee}\left(q\right) & = & \chi_{0}^{\textrm{\tiny SU(2)}_{3}}\left[\chi_{\mathbb{I}}^{\mathbb{Z}_{3}}+\chi_{\zeta}^{\mathbb{Z}_{3}}+\chi_{\zeta^{*}}^{\mathbb{Z}_{3}}\right]+\chi_{1}^{\textrm{\tiny SU(2)}_{3}}\left[\chi_{\Psi}^{\mathbb{Z}_{3}}+\chi_{\Psi^{*}}^{\mathbb{Z}_{3}}+\chi_{\Omega}^{\mathbb{Z}_{3}}\right],\nonumber \\
Z_{OO}^{ooo}\left(q\right) & = & \chi_{3/2}^{\textrm{\tiny SU(2)}_{3}}\left[\chi_{\mathbb{I}}^{\mathbb{Z}_{3}}+\chi_{\zeta}^{\mathbb{Z}_{3}}+\chi_{\zeta^{*}}^{\mathbb{Z}_{3}}\right]
+\chi_{1/2}^{\textrm{\tiny SU(2)}_{3}}\left[\chi_{\Psi}^{\mathbb{Z}_{3}}+\chi_{\Psi^{*}}^{\mathbb{Z}_{3}}+\chi_{\Omega}^{\mathbb{Z}_{3}}\right].  \qquad
\label{eq:ZOO}
\end{eqnarray}
Here  $\Psi$ and $\Psi^{*}$ denote  the pair of  conjugate fields of dimension $3/5$ (see \ref{app:Z3}) and $\Omega$ is 
the dimension-$8/5$ field generated from their fusion. The conjugate  fields $\zeta$
and $\zeta^{*}$ have dimension 2.  Explicit expression for the characters are given in \cite{di1997conformal,Totsuka1996,frenkel1992}.

We can read off the finite-size spectrum for three decoupled spin chains from the partition functions in (\ref{eq:ZOO}) \cite{di1997conformal}.  
In particular, the ground state for three even chains is a singlet, and all the conformal towers in $Z_{OO}^{eee}$ are associated with operators of 
integer scaling dimension. By contrast, the ground state for three odd chains is eightfold degenerate, as expected from the
direct product of three spin doublets, cf.~equation (\ref{eq:ZOOaff}).
Furthermore, we may combine equation (\ref{eq:ZOO}) with the physical intuition gained from the bosonization approach to analyze the 
symmetry properties of operators in the $\mathbb Z_3^{(5)}$ sector, as in \cite{Ludwig1994} for the multi-channel Kondo model. 
Let us consider the partition function of three even chains, $Z^{eee}_{OO}$.
There are nine fields with dimension 1 in both formulations  (\ref{eq:ZOOaff}) and (\ref{eq:ZOO}).
In equation (\ref{eq:ZOOaff}), these can be identified with the chiral currents $J^a_{\alpha,L}$, with
$\alpha=1,2,3$ and $a=1,2,3$. In order to identify the dimension-1 operators in the  SU(2)$_3\times\mathbb Z_3^{(5)}$ embedding, it is convenient to 
introduce the helical combinations \cite{Ingersent2005}
\begin{eqnarray}
 \mathcal{J}^a_{h}(z)&=&\sum_{\alpha=1}^{3}\omega^{-\alpha h} {J}^a_{\alpha,L}(z), 
  \qquad\qquad h=-1,0,1.\label{helicalcurrent}
\end{eqnarray}
Clearly, $ \mathcal{J}^a_{0}$ coincides with  the SU(2)$_3$  current $\mc J^a$ defined in equation (\ref{J0current}). The combinations $ \mathcal{J}^a_{\pm1}$ transform as $ \mathcal{J}^a_{\pm1}\mapsto  \omega^{\pm1}\mathcal{J}^a_{\pm1}$ under the cyclic permutation $J^a_{\alpha,L} \mapsto J^a_{\alpha+1,L}$.
  Considering the Kac-Moody algebra (\ref{KMope}), the helical currents obey the OPE 
\begin{equation}
\mathbf{\mathcal{J}}_{h}^a
\left(z\right)\mathbf{\mathcal{J}}^b_{h'}\left(w\right)\sim\frac{3\delta_{hh'}\delta_{ab}}{8\pi^2\left(z-w\right)^{2}}+\frac{i\varepsilon_{abc}}{2 \pi (z-w)}\mathbf{\mathcal{J}}^c_{h+h'}\left(w\right),\label{HelicalCurrentOPE}
\end{equation}
where the sum $h+h'$ is defined modulo $3$ and restricted to the domain
$\{-1,0,1\}$.  Having identified   $\mc J_0^a$ with the SU(2)$_3$ current, we are left with the six components $\mc J_{\pm1}^a$.  
The only other dimension-1 fields in the partition function (\ref{eq:ZOO}) are the products
$\phi_1^{a} \Psi$ and $\phi_1^{a}\Psi^{*}$. Here ${\bm\phi}_1$ is the spin-1 primary field (the vector) of the SU(2)$_3$ WZW model, which has scaling dimension $2/5$, 
while $\Psi$ and $\Psi^*$ are conjugate fields of dimension $3/5$. The helical currents $\mc J_{\pm1}^a$ must   be linear combinations
of $\phi_1^{a} \Psi$ and $\phi_1^{a}\Psi^{*}$ chosen such that the OPE (\ref{HelicalCurrentOPE}) is compatible with the fusion algebra for $su(2)_3$ and $\mathbb Z_3^{(5)}$. 
We then consider linear combinations with coefficients $c_\alpha$ and $d_\alpha$,
\begin{eqnarray}
 &  & \phi_1^{a}\Psi\sim\sum_{\alpha}c_{\alpha}J_{\alpha}^{a}+\cdots,\qquad\qquad\phi_1^{a}\Psi^{*}\sim\sum_{\alpha}d_{\alpha}J_{\alpha}^{a}+\cdots,\label{Phipsi-moreorless}
\end{eqnarray}
where $\cdots$ stands for operators with higher conformal dimension.  Using the OPE for  level-$1$ currents in  (\ref{KMope}), we find
\begin{eqnarray}
\sum_{\alpha,\beta}c_{\alpha}d_{\beta}J_{\alpha}^{a}\left(z\right)J_{\beta}^{b}\left(w\right) & \sim &
\frac{\left(\sum_{\alpha}c_{\alpha}d_{\alpha}\right)\mathbb{I}}{\left(z-w\right)^{2}}+\frac{i\varepsilon_{abc}\sum_{\alpha}c_{\alpha}d_{\alpha}J_{\alpha}^{c}\left(w\right)}{z-w},\nonumber \\
\sum_{\alpha,\beta}c_{\alpha}c_{\beta}J_{\alpha}^{a}\left(z\right)J_{\beta}^{b}\left(w\right) & \sim &
\frac{\left(\sum_{\alpha}c_{\alpha}^{2}\right)\mathbb{I}}{\left(z-w\right)^{2}}+\frac{i\varepsilon_{abc}\sum_{\alpha}c_{\alpha}^{2}J_{\alpha}^{c}\left(w\right)}{z-w},
\end{eqnarray}
for the divergent part. 
On the other hand, the fusion rules calculated from the left hand side of (\ref{Phipsi-moreorless})
have to be satisfied, in particular $\Psi\times\Psi\sim\Psi^{*}$ plus a regular part, 
see table \ref{tab:Z3-FusionRules} and (\ref{su(2)fusionCoefficients}).
This is   sufficient to fix $c_{\alpha}=
\omega^{-\alpha}=d_{\alpha}^{*}$,
up to an overall phase $\omega$ or $\omega^\ast$ (corresponding to a $\mathbb{Z}_{3}$ permutation) and complex conjugation (corresponding
to the exchange $\Psi\leftrightarrow\Psi^{*}$ or $1\leftrightarrow 2$ for leg indices).
Summing up, we conjecture that the helical combinations can be identified as 
\begin{equation}
\left(\begin{array}{c}
\mc J_0^a\\
\mc J_1^a\\
\mc J_{-1}^a
\end{array}\right)=
\left(\begin{array}{c}
\mathcal{J}^{a}\\
\phi^a_1\Psi\\
\phi^{a}_1\Psi^{*}
\end{array}\right)=\left(\begin{array}{ccc}
1 & 1 & 1\\
\omega^* & \omega & 1\\
\omega & \omega^* & 1
\end{array}\right)\left(\begin{array}{c}
J_{1,L}^{a}\\
J_{2,L}^{a}\\
J_{3,L}^{a}
\end{array}\right)\label{eq:Currents-embedding-rotation} .
\end{equation}
A more careful determination of the OPE coefficients in this conformal embedding passes through the evaluation of four-point functions \cite{Ludwig1994}, but
we do not tackle this problem here. The important point is that the transformation of the $\Psi$ and $\Psi^*$ fields in the $\mathbb Z_3^{(5)}$ sector are the same as for the helical currents $\mc J_{\pm1}^a$ under cyclic chain permutation. In particular, they transform under $\alpha\mapsto \alpha+1$ as
\be
\Psi \mapsto \omega\Psi,\qquad\qquad \Psi^*\mapsto \omega^*\Psi^*.
\ee
Since both $\mc P$ and $\mc T$ exchange the helical currents $\mc J_{1}^a\leftrightarrow \mc J_{-1}^a$, these transformations must  act in the $\mathbb Z_3^{(5)}$ sector by exchanging $\Psi\leftrightarrow \Psi^*$. 
We can also deduce from the fusion rules that the dimension-$8/5$ operator $\Omega$ is invariant under cyclic chain permutation but odd under $\mc P$ and $\mc T$ (see \cite{Buccheri2018} and section \ref{subsec:Spin-chirality-operator}). 
 
 We now apply the fusion procedure to generate other conformally invariant boundary conditions. Let us first consider the K fixed point, at which the three chains overscreen  the effective spin-$1/2$ degree of freedom resulting from the strongly coupled end spins. Under the fusion hypothesis \cite{Affleck:1990zd,Affleck1991b},
the spectrum of the three-channel Kondo fixed point is obtained by fusing the partition function for open boundary conditions with the character
of the spin-$1/2$ field in the  SU(2)$_3$ sector. The fusion rules are recalled in (\ref{su(2)fusionCoefficients}) in \ref{app:WZW}. 
 The partition function with the same boundary condition at both ends is generated by double fusion \cite{Affleck1995}. For even-length chains, we obtain
\begin{eqnarray}
Z^{eee}_{KK}\left(q\right) & = & \chi_{0}^{\textrm{\tiny SU(2)}_{3}}\left[\chi_{\mathbb{I}}^{\mathbb{Z}_3}
+\chi_{\zeta}^{\mathbb{Z}_3}+\chi_{\zeta^{*}}^{\mathbb{Z}_3}
+\chi_{\Psi}^{\mathbb{Z}_3}+\chi_{\Psi^{*}}^{\mathbb{Z}_3}
+\chi_{\Omega}^{\mathbb{Z}_3}\right]\nonumber \\
 &  & +\chi_{1}^{\textrm{\tiny SU(2)}_{3}}\left[\chi_{\mathbb{I}}^{\mathbb{Z}_3}
 +\chi_{\zeta}^{\mathbb{Z}_3}+\chi_{\zeta^{*}}^{\mathbb{Z}_3}
 +2\chi_{\Psi}^{\mathbb{Z}_3}+2\chi_{\Psi^{*}}^{\mathbb{Z}_3}
 +2\chi_{\Omega}^{\mathbb{Z}_3}\right].\label{Z_KK}
\end{eqnarray}
Equation (\ref{Z_KK}) allows us to identify the   boundary operators in the effective Hamiltonian for  the K fixed point. These operators must be   SU(2) scalars and invariant under $\mathbb Z_3$ and $\mc{PT}$ symmetries. Thus,  we can rule out the relevant (dimension $3/5$) operators  $\Psi$ and $\Psi^*$, which are not invariant under cyclic permutations of the chains. It turns out that all the boundary  operators  allowed by symmetry are irrelevant, implying that the K fixed point is stable and can thus describe the low-energy physics of the Y junction in the regime $|J_\chi|\gg J_1$.  This fusion hypothesis must be tested by comparing the analytical predictions against numerical results, as we shall do in section \ref{sec:Numerical-checks}.  As in the free-electron multi-channel Kondo model \cite{Affleck1991}, the leading irrelevant operator is the  Kac-Moody descendant of the spin-1 field,  ${\bm{\mc J}}_{-1}\cdot {\bm\phi}_1$. Here ${\bm{\mc J}}_{n}$ denotes the modes of the SU(2)$_3$ chiral current  in the Laurent expansion, ${\mc J}^a(z)=\sum_{n\in \mathbb Z} z^{-n-1}  {\mc J}^a_n$ \cite{di1997conformal}. This boundary operator has   dimension $7/5$ and is time-reversal invariant. The leading operator which is odd under $\mc P$ and $\mc T$ is the dimension-$8/5$ field $\Omega$ acting in the $\mathbb Z_3^{(5)}$ sector. Note that this operator does not occur in the free-electron three-channel Kondo model, where the flavor sector is represented by an SU$(3)_2$ WZW theory \cite{Affleck1991}. As a result, the leading perturbations to the Kondo fixed point are given by
\be
H_B^{(K)}=\lambda_1^{(K)} {\bm{\mc J}}_{-1}\cdot {\bm\phi}_1+\lambda_2^{(K)}\Omega+\cdots . \label{HBK}
\ee

Finally, we discuss how to obtain the chiral fixed points C$_\pm$ using fusion.  Since these fixed points break the leg exchange symmetry $\mc P$, we expect them to be generated from the open boundary partition function by fusion with judiciously  chosen primary fields in the  $\mathbb Z_3^{(5)}$ sector.  The criteria to guide our choice of  the fields are as follows. (i) The two chiral fixed points with opposite chirality should be associated with a pair of conjugate primary fields with the same dimension. (ii) The spectrum must reproduce the result for three chiral SU(2)$_1$ models since the boundary conditions can, in the bosonization approach, be described by (\ref{bcchiralJ}).   (iii) The leading boundary operator allowed by symmetry must be a scalar with scaling dimension $1/2$.   (iv) The spin conductance matrix calculated from the corresponding  boundary states (see section \ref{subsec:spin-conductance}) must be asymmetric and reflect the transmission of spin currents according to the boundary conditions (\ref{bcchiralJ}). These criteria lead us to postulate that the chiral fixed points are obtained by fusion with the dimension-$1/9$ fields denoted by $\xi$ and $\xi^*$ in \ref{app:Z3}.  Within double fusion, we must fuse first with one field (representing the first boundary in the strip geometry) and second with the conjugate field (representing the mirror image at the other boundary). The order of the fusion, $\xi\xi^*$ or $\xi^*\xi$, selects the boundary state corresponding to either  C$_+$ or C$_-$, but the partition function is the same in both cases,   
\begin{eqnarray}
Z^{eee}_{CC}\left(q\right) & = & 
\chi_{0}^{\textrm{\tiny SU(2)}_{3}}\left[\chi_{\mathbb{I}}^{\mathbb{Z}_{3}}+3\chi_{\varepsilon'}^{\mathbb{Z}_{3}} 
+\chi_{\zeta}^{\mathbb{Z}_{3}}
+\chi_{\zeta^{*}}^{\mathbb{Z}_{3}}\right]\label{Zchichi*}\nonumber \\
 &  & +\chi_{1}^{\textrm{\tiny SU(2)}_{3}}\left[3\chi_{\varepsilon}^{\mathbb{Z}_{3}}
 +\chi_{\Psi}^{\mathbb{Z}_{3}}+\chi_{\Psi^{*}}^{\mathbb{Z}_{3}}
 +\chi_{\Omega}^{\mathbb{Z}_{3}}\right].\label{chiralZ}
\end{eqnarray}
Here $\varepsilon$ and $\varepsilon'$ are the primary fields with dimension $1/10$ and $1/2$, respectively.  Interestingly, the spectrum contains three towers of states associated with $\varepsilon'$, where the $\mathbb Z_3$ symmetric combination of such states corresponds to the backscattering  operator 
$\sum_\alpha \textrm{Tr}[\tilde g_\alpha]$. 
Note that there are, in addition, three towers associated with the spinful operator ${\bm\phi}_1\varepsilon$, also with dimension $1/2$. This operator can appear in the effective Hamiltonian for the chiral fixed point if combined with a free boundary spin, as in $\mb S_0\cdot{\bm\phi}_1\varepsilon$. In the bosonization approach, this corresponds to the boundary operator $\mb S_0\cdot \sum_\alpha \textrm{Tr}[\tilde g_\alpha {\bm\sigma}]$. In the strong coupling  regime $|J_\chi|>J_\chi^c$, this operator destabilizes the chiral fixed points, signalling the RG flow towards the K fixed point where the boundary spin gets screened. Except for these two dimension-$1/2$ operators and their descendants, all other states in the partition function (\ref{chiralZ}) come with integer scaling dimension as expected for  chiral SU(2)$_1$ WZW models. 

\subsection{Boundary entropy}

As a consequence of nontrivial boundary conditions, 1D
quantum systems can acquire a universal
non-integer ground-state degeneracy, $g$, in the thermodynamic limit. The Affleck-Ludwig boundary entropy \cite{Affleck1991b} appears as a non-extensive
correction to the free energy for $\mc L\gg v\beta$,
\be
\ln Z=\frac{\pi c}{6}\frac{\mc L}{v\beta} +\ln g+\cdots,
\ee
where $c$ is the central charge.
For multi-channel Kondo fixed points, the boundary entropy can be  computed from the modular $\mathcal{S}$ matrix 
of the SU(2)$_{k}$ WZW theory  \cite{Affleck1991b,di1997conformal}. The components of the $\mc S$ matrix are labeled by the spin of the primary, $j_1,j_2=0,\frac12,1,\frac32$, and are given by
\begin{eqnarray}
\mathcal{S}_{j_{1}j_{2}}^{\textrm{\tiny SU(2)}_{3}} & = & \sqrt{\frac{2}{5}}\sin\left(\frac{\pi\left(2j_{1}+1\right)\left(2j_{2}+1\right)}{5}\right).\label{eq:modularS-su(2)_3}
\end{eqnarray}
The ratio of the ground-state degeneracies at the K and O fixed points follows as
\begin{equation}
\frac{g_{K}}{g_{O}}=\frac{\mathcal{S}_{\frac{1}{2},0}^{\textrm{\tiny SU(2)}_{3}}}{\mathcal{S}_{0,0}^{\textrm{\tiny SU(2)}_{3}}}=2\cos\frac{\pi}{5}\approx 1.62 .\label{gKondo}
\end{equation}
For the chiral fixed points, we use the $\mc S$ matrix for the  $\mathbb{Z}_{3}^{(5)}$ theory \cite{Fateev1990,frenkel1992},
see \ref{app:Z3}. The ratio
between the ground-state  degeneracy of the C$_\pm$ and O fixed points reads
\begin{equation}
\frac{g_{C_+}}{g_{O}}=\frac{g_{C_-}}{g_{O}}=\frac{\mathcal{S}_{\xi,\mathbb{I}}^{\mathbb{Z}_{3}}}{\mathcal{S}_{\mathbb{I},\mathbb{I}}^{\mathbb{Z}_{3}}}=2 . \label{gChiral}
\end{equation}
According to the $g$-theorem \cite{Affleck1991b,Friedan2004}, the ground-state degeneracy can only decrease under the RG flow. Our boundary phase diagram in figure \ref{fig2} is consistent with the $g$-theorem since  the chiral fixed points C$_\pm$ have the highest $g$ value, indicating that boundary perturbations trigger an RG flow towards either the O or the K fixed point. 
 
\section{Characterization of the fixed points\label{sec:FixedPoints}}

We now study physical observables that can be used in numerical or experimental tests of our proposed scenario.

\subsection{Scalar spin chirality \label{subsec:Spin-chirality-operator}}

Since it breaks time-reversal symmetry, a nonzero scalar spin chirality   can in principle be probed  by circular dichroism \cite{Kitamura2017,Bulaevskii2008}. The expectation value of the  SSCO  also provides a numerical test of the boundary phase diagram, since the  decay of the three-spin correlation function at large distances from the boundary is governed by  the fixed point in each regime of $J_\chi$ \cite{Buccheri2018}. 
To discuss three-spin correlations, let us first rewrite the SSCO  (\ref{SSCO}) in the continuum limit using the conformal embedding. The term with the lowest bulk scaling dimension stems from the staggered magnetization for each chain,
\bea
\hat C(\tau,x)&\sim&\left(-1\right)^{x}\varepsilon_{abc}n^a_1(\tau,x)n^b_2(\tau,x)n^c_3(\tau,x)\nonumber\\
&=& \left(-1\right)^{x}\mathcal{A}^{3}\varepsilon_{abc}\mbox{Tr}\left[g_{1}(\tau,x)\sigma^{a}\right]\mbox{Tr}\left[g_{2}(\tau,x)\sigma^{b}\right]\mbox{Tr}\left[g_{3}(\tau,x)\sigma^{c}\right],\label{CnnnNab}
\eea
where all fields are  evolved in imaginary time.
The  operator in equation (\ref{CnnnNab}) is an SU(2) scalar with zero conformal spin and  scaling dimension $3/2$. Moreover, it is odd under $\mc P$ and $\mc T$ but invariant under $\mathbb Z_3$ cyclic chain permutation. Using these properties, we select its counterpart  in the SU(2)$_3\times \mathbb Z_3$ formulation,
\begin{eqnarray}
\hat C \left(\tau,x\right) & \sim &(-1)^x
 i\mbox{Tr}\left[\phi_{\frac12}\left(z\right)\otimes \phi^\dagger_{\frac12}\left(\bar{z}\right)\right]
\left[\Psi\left(z\right)\Psi^{*}\left(\bar{z}\right)
-\Psi^{*}\left(z\right)\Psi\left(\bar{z}\right)\right],
\label{Z3Cnnn}
\end{eqnarray}
where $\phi_{\frac12}$ is the spin-$\frac{1}{2}$ primary field of the SU(2)$_3$ WZW model and $\phi_{\frac12}\left(z\right)\otimes \phi^\dagger_{\frac12}\left(\bar{z}\right)$ is the corresponding matrix field. 
Note  the conformal dimensions $\Delta=\bar{\Delta}=\frac{3}{20}+\frac{3}{5}=\frac{3}{4}$. 

For open boundary conditions, the boundary spins reduce  to  chiral currents,  $\mb S_{1,\alpha}\propto \mb J_{\alpha,L}(0)$, and the boundary SSCO  becomes proportional to the triple product $\hat C(x=0)\propto \varepsilon_{abc}J^a_{1,L}(0)J^b_{2,L}(0)J^c_{3,L}(0)$.  The corresponding operator, which appears in the partition function (\ref{eq:ZOO}) and is generated from the OPEs of the  fields  in equation (\ref{Z3Cnnn}) \cite{Zamolodchikov1986,Lukyanov1988,Fateev1987,Zamolodchikov1985} in the boundary limit $\bar z\to z\to v \tau$,
is given by
\begin{equation}
\hat C(\tau,x=0)\sim\boldsymbol{\mathcal{J}}_{-1}\cdot\boldsymbol{\phi}_1\Omega\qquad\qquad (\textrm{O fixed point}).\label{CJJJnab}
\end{equation}
At the three-channel Kondo fixed point, the partition function (\ref{Z_KK}) contains the boundary operator $\Omega$, which has the same symmetries as the SSCO. This operator is obtained from equation (\ref{Z3Cnnn}) if the spin-$\frac{1}{2}$ fields are allowed to fuse to the identity at the boundary. Therefore, at the K fixed point the boundary SSCO is represented by the dimension $8/5$ field
\be
\hat C(\tau,x=0)\sim\Omega\qquad\qquad (\textrm{K fixed point}).
\ee

Let us now discuss the large-distance decay of the three-spin correlation \cite{Buccheri2018},
\be
G_3(x)=\langle \hat C_{j=x}\rangle.
\ee
In the BCFT approach, this amounts to the calculation of a one-point function for given boundary conditions.  
At the chiral fixed point, the SSCO has a nonzero expectation
value. From equation (\ref{Z3Cnnn}), we need to evaluate
\be
\left \langle \phi^{\sigma}_{\frac12}(z=ix) \phi^{\dagger\sigma'}_{\frac12}(\bar z=-ix)\right\rangle_C \sim \delta_{\sigma\sigma'}x^{-\frac3{10}},\label{halfhalf}
\ee
where $\sigma,\sigma'=1,2$ label the components of the fundamental spinor. We also need
\begin{eqnarray}
i\left[\left\langle \Psi\left(ix\right)\Psi^{*}\left(-ix\right)\right\rangle _C-\left\langle \Psi\left(ix\right)\Psi^{*}\left(-ix\right)\right\rangle _C \right]& = & \frac{\textrm{Im}\mathcal{B}_{\Psi}^{\xi}}{x^{6/5}},\label{PsiPsif}
\end{eqnarray}
where the modular $\mc S$ matrix in \ref{app:Z3} is used to compute 
\begin{equation}
\mathcal{B}_{\Psi}^{\xi}=\frac{\mathcal{S}_{\xi,\Psi}^{\mathbb{Z}_{3}}\mathcal{S}_{\mathbb{I},\mathbb{I}}^{\mathbb{Z}_{3}}}{\mathcal{S}_{\mathbb{I},\Psi}^{\mathbb{Z}_{3}}\mathcal{S}_{\xi,\mathbb{I}}^{\mathbb{Z}_{3}}}=\omega^*.\label{BoundaryBchiPsi}
\end{equation}
Combining equations (\ref{halfhalf}) and (\ref{PsiPsif}), we  obtain the decay
\begin{equation}
G^{(C)}_3(x)\sim\left(-1\right)^{x}x^{-\frac{3}{2}},\label{CnnDecayChi}
\end{equation}
which holds for both C$_+$ and C$_-$ fixed points. We can obtain the same result  from bosonization. Using the staggered part of the spin operators and the  boundary conditions (\ref{bcbosons}), the three-spin correlation at the chiral fixed point can be written as 
\begin{equation}
G^{(C)}_{3}(x)\sim  (-1)^x \prod_{\alpha=1}^3
 \left\langle e^{i\sqrt{\pi}\varphi_{\alpha,L}(x)}e^{-i\sqrt{\pi}\varphi_{\alpha,L}(-x)}\right\rangle\sim  (-1)^xx^{-\frac32}.
\end{equation}

Turning next to the time-reversal symmetric O and K fixed points, we first note that the SSCO expectation values vanish identically right at these points.
 To obtain a nonzero result, we apply perturbation theory in the leading irrelevant boundary operators,
\bea
G^{(O,K)}_3(x)&=&\left\langle \hat C(x) e^{-\int_{-\infty}^{\infty} d\tau\, H^{(O,K)}_B(\tau)}\right\rangle_{O,K} \nonumber\\
&\sim &-\int_{-\infty}^{\infty} d\tau\, \left\langle \hat C(x) H^{(O,K)}_B(\tau)\right\rangle_{O,K} .
\eea
In the boundary Hamiltonian $H_B$, we select the leading operator with the same symmetries as the SSCO.  

For the O fixed point, this is the operator $\sim\lambda_3^{(O)}$ in equation (\ref{HBO}). Using bosonization, we  have
\bea
 G^{(O)}_3(x)
&\sim&\lambda_3^{(O)}(-1)^x\int_{-\infty}^{\infty} d\tau\, \varepsilon_{abc} \varepsilon_{a'b'c'}\langle n_1^a(x)J^{a'}_{1,L}(\tau)  \rangle \nonumber\\
&&\times\langle n_2^b(x)J^{b'}_{2,L}(\tau)  \rangle \langle n_3^c(x)J^{c'}_{3,L}(\tau)  \rangle \nonumber\\
&&\propto (-1)^xx^{-7/2}.\label{G3O}
\eea 
Here we used that the chiral current $\mathbf J_{\alpha,L}$ and the staggered magnetization $\mathbf n_\alpha$ can fuse to the dimerization operator, which has a nonzero expectation value in the open chain. On the other hand, using the conformal embedding, we can write the boundary SSCO as in equation (\ref{CJJJnab}). The correlation function then factorizes  into   SU(2)$_{3}$
and $\mathbb{Z}_{3}^{(5)}$ sectors,
\begin{eqnarray}
G_3^{(O)}(x)& \sim &  (-1)^x\int_{-\infty}^{\infty} d\tau\, \left\langle \mathcal{J}_{-1}^{a}\mathbf{\phi}_1^{a}\left(\tau\right)\phi_{\frac12}^{\sigma}\left(ix\right)\phi_{\frac12}^{\dagger\sigma}\left(-ix\right)\right\rangle_O\nonumber
\\&&
\times i \left[\left\langle \Omega\left(\tau\right)\Psi\left(ix\right)\Psi^{*}\left(-ix\right)\right\rangle_O -\left\langle \Omega\left(\tau\right)\Psi^{*}\left(ix\right)\Psi\left(-ix\right)\right\rangle_O \right],
\end{eqnarray}
which can be evaluated using the fusion rules and again yields (\ref{G3O}). 

Near the K fixed point, the leading contribution originates from the term $\sim \lambda_2^{(K)}$ in (\ref{HBK}). We obtain
\begin{eqnarray}
G_3^{(K)}(x)
& \sim &  (-1)^x\int_{-\infty}^{\infty} d\tau\, \left\langle\phi_{\frac12}^{\sigma}\left(ix\right)\phi_{\frac12}^{\dagger\sigma}\left(-ix\right)\right\rangle_K\nonumber
\\&&
\times i \left[\left\langle \Omega\left(\tau\right)\Psi\left(ix\right)\Psi^{*}\left(-ix\right)\right\rangle_O -\left\langle \Omega\left(\tau\right)\Psi^{*}\left(ix\right)\Psi\left(-ix\right)\right\rangle_K \right]
 \nonumber\\
&\propto& (-1)^xx^{-21/10}.
\end{eqnarray}

In summary, the  three-spin correlation function has the asymptotic ($x\to \infty$) power-law  decay
\be
G_3(x)\sim (-1)^xx^{-\nu},\label{G3oscdecay}
\ee
where the exponent is characteristic for the respective fixed point,
\be
\nu_O=\frac72,\qquad\qquad \nu_C= \frac32,\qquad\qquad\nu_K=\frac{21}{10}.\label{G3exponents}
\ee
Note that the chiral fixed point has the smallest exponent corresponding to the slowest decay of $G_3(x)$.
 
\subsection{Spin conductance\label{subsec:spin-conductance}}

Recent experiments have shown that   antiferromagnets  can act as efficient conductors of 
spin currents, essentially without
involving charge transport \cite{Lebrun2018}. Here, we envision a setup in which a spin current  is injected from a metal  into a spin chain that forms one of the legs of a Y junction. The spin chain could be realized, for instance, by arranging spin-$\frac{1}{2}$ atoms on a surface using a scanning tunnelling microscope (STM) \cite{Toskovic2016,Yang2017}.  The spin current could be generated by spin accumulation due to the spin Hall effect in the metal.  The difference in chemical potential for spin-up and spin-down electrons in
the  terminals   plays the role of a 
magnetic field  $B^z_\alpha=\mu_{\alpha,\uparrow}-\mu_{\alpha,\downarrow}$ at the end of chain $\alpha=1,2,3$. We note in passing that the spin chemical potential of magnons has recently been measured with high resolution \cite{Du2017}. Alternatively,  $B^z_\alpha$ could represent  the external magnetic field  of a  magnetic STM tip. The field can be oriented in any   direction by suitably modifying the setup. While the charge of the electrons cannot propagate into the antiferromagnetic insulator at low energies, the gradient of magnetic field at the metal-insulator interface drives a spin current into the spin chain. In this case, the elementary spin-carrying excitations are the spinons of the Heisenberg chain, as opposed to magnons in an ordered antiferromagnet. The spin current transmitted to the other legs of the Y junction could  then be detected by converting it back to a charge current in the attached metallic terminal via the inverse spin Hall effect. 

One may exploit the similarities  to charge transport in quantum wires to define a spin conductance for spin-$1/2$ chains 
 \cite{Meier2003}. Let $I_\alpha^{a}$ denote the spin current component polarized along direction $a$ flowing into the junction from chain $\alpha$. Within linear response theory, 
 spin transport through the Y junction is then characterized by a spin conductance tensor $G$,
\begin{equation}
I_{\alpha}^{a}=\sum_{b,\beta} {G}_{\alpha\beta}^{ab}B_{\beta}^{b},\label{spinGdef}
\end{equation}
where $B_\beta^b$ is the magnetic field or, more precisely, the spin chemical potential along direction $b$ at the end of chain $\beta$. In the presence of SU(2) symmetry, the spin current is parallel to the field that drives it. The conductance tensor must therefore be diagonal in the spin indices, 
\be
{G}_{\alpha\beta}^{ab}=\delta_{ab}G_{\alpha\beta}. 
\ee 
In analogy with  charge conservation in quantum wires \cite{Oshikawa2006}, total spin conservation in the junction implies, for arbitrary spin chemical potentials,  the Kirchhoff node rule, $\sum_\alpha I^a_\alpha=0$. This implies the constraint $\sum_{\alpha}{G}_{\alpha\beta}=0$. Moreover, since spin currents only flow when there is a spin chemical potential difference between the terminals, we must have $I^a_\alpha=0$ if $\mathbf B_\beta=\mathbf B$ is identical for all chains. As a result, we also have  $\sum_{\beta}{G}_{\alpha\beta}=0$. Finally imposing the $\mathbb{Z}_3$ symmetry of our Y junction, the general form of the linear spin conductance tensor compatible with all constraints is 
\begin{equation}
 {G}_{\alpha\beta}^{ab}=\frac{1}{2}\delta_{ab}\left[G_{S}\left(3\delta_{\alpha\beta}-1\right)+G_{A}\varepsilon_{\alpha\beta}\right],
\label{Z3conductance}
\end{equation}
where $\varepsilon_{\alpha\beta}$ is the Levi-Civita symbol with $\varepsilon_{12}=\varepsilon_{23}=\varepsilon_{31}=1$ and $\varepsilon_{\alpha\beta}=-\varepsilon_{\beta\alpha}$. The two parameters $G_{S}$ and $G_A$ characterize the symmetric and antisymmetric parts of the spin conductance tensor, respectively. 
Below we determine their values at the various fixed points of the Y junction model.
  
\subsubsection{Hydrodynamic approach at the chiral fixed point.}

Before tackling a more formal derivation, we  provide an intuitive hydrodynamic
approach to the problem of computing the spin conductance at the chiral fixed point.
Following  \cite{Balents2001}, we rewrite the Heisenberg equations of motion  for the chiral currents $\mathbf J_{\alpha,L}$ and $\mathbf J_{\alpha,R}$ in each chain    as coupled equations for the magnetization,
\be
\mb M_\alpha(x)=\mb J_{\alpha,R}(x)+\mb J_{\alpha,L}(x),\label{magnop}
\ee
and for the spin current,
\be
\mb J_\alpha(x)=v[\mb J_{\alpha,R}(x)-\mb J_{\alpha,L}(x)].\label{spincurrop}
\ee
Here we have set $\gamma=0$ in equation  (\ref{continuumlimitchain}) to neglect  the marginally irrelevant bulk operator. We then obtain two equations by taking the sum and the difference of the Heisenberg equations. The first equation is simply the spin continuity equation,
\begin{equation}\label{spincont}
\partial_t {\bf M}_\alpha + \partial_x {\bf J}_\alpha = 0 .
\end{equation}
The second equation is   
\begin{equation}\label{magn}
\partial_t {\bf J}_\alpha+v^2 \partial_x {\bf M}_\alpha=0.
\end{equation}
Within the hydrodynamic approach, one replaces the operators $\mb M_{\alpha}(x)$ and $\mb J_\alpha(x)$ by their `classical' expectation values \cite{Balents2001}. In the steady state, $\partial_t {\bf M}_\alpha \to 0$, the continuity equation implies that the spin current is uniform  in each chain, 
${\bf J}_\alpha (x) =\mb  I_\alpha.$ 
The second equation implies that the magnetization $\mb M_\alpha$ is also constant, varying only near the contacts.    
When different chains are coupled by very weak tunnelling processes, i.e., we are near
the O fixed  point,
the nonequilibrium distribution function in each chain can be described 
by a spin chemical potential vector ${\bf B}_\alpha$, where ${\bf M}_\alpha=\chi_0
{\bf B}_\alpha$ with the spin susceptibility $\chi_0=1/(2\pi v)$. 
Note that ${\bf B}_\alpha$ does not represent a magnetic field but characterizes only
 the distribution function due to attached spin reservoir \cite{Balents2001}.   
However, near the 
C$_\pm$ fixed points, we should proceed in a different manner. Suppose that $H_B$ realizes ideal chiral boundary conditions, say C$_-$ with
${\bf J}_{\alpha,R}(0)={\bf J}_{\alpha+1,L}(0)$.
 We find  
\begin{equation}
{\bf J}_{\alpha+1}+{\bf J}_{\alpha}=v\left({\bf M}_{\alpha+1}-{\bf M}_\alpha
\right).
\end{equation} 
This in turn is consistent with the Kirchhoff node rule, $\sum_\alpha {\bf J}_\alpha=0$, 
and we get
\begin{equation}
{\bf J}_{\alpha}=  v  \left({\bf M}_{\alpha+1}-{\bf M}_{\alpha-1}
\right).
\end{equation}
The spin chemical potentials here regulate only the incoming spin densities,
${\bf J}_{\alpha,L}=\chi_0{\bf B}_{\alpha}$.
The spin currents are then given by
\begin{equation}
{\bf J}_\alpha = v({\bf J}_{\alpha,R}-{\bf J}_{\alpha,L})=
v({\bf J}_{\alpha+1,L}-{\bf J}_{\alpha,L}).
\end{equation}
Using the $x$-independence of the spin currents, $\mb J_\alpha=\mb I_\alpha$, we get
\begin{equation}
{\bf I}_\alpha = \chi_0 v({\bf B}_{\alpha+1}-{\bf B}_\alpha).
\end{equation}
Assuming that all spin chemical potentials and spin currents are taken along
the $z$-axis, we have
\begin{equation}
 {G}_{\alpha\alpha'} = -\frac{\partial J^z_\alpha}{\partial B_{\alpha'}}
= \frac{1}{2\pi} \left(\delta_{\alpha',\alpha}-\delta_{\alpha',\alpha+1}\right).\label{conductancehydro}
\end{equation}
We can write this in the form of equation  (\ref{Z3conductance}) with $G_S=G_A=1/(2\pi)$.   The associated  spin conductance tensor   is maximally asymmetric  in the sense that $G_{\alpha-1,\alpha}=-1/(2\pi)$ while $G_{\alpha+1,\alpha}=0$. Note that the magnitude of $G_{\alpha-1,\alpha}$   equals   the quantum of spin conductance, 
\be
G_0=\frac1{2\pi},\label{conductancequantum}
\ee
in units where $g\mu_B=\hbar=1$ \cite{Meier2003}.  This means that  the spin current injected into chain $\alpha$ is fully transmitted in a clockwise rotation to chain $\alpha-1$, thus realizing an ideal spin circulator \cite{Buccheri2018}. 

\subsubsection{Kubo formula.\label{sec:KuboCFT}}

The linear spin conductance (taken at zero temperature)  can alternatively be computed using the Kubo formula \cite{Meier2003,Oshikawa2006},
\begin{eqnarray}
 {G}_{\alpha\beta}^{ab} & = & -\lim_{\omega\to0^+} \frac{1}{\omega \mc L}\intop_{0}^{\mc L}dx\intop_{-\infty}^{\infty}d\tau\,e^{i\omega\tau}{\cal G}_{\alpha,\beta}^{a,b}\left(x,y;\tau\right)\label{eq:Kubo},
\end{eqnarray}
with chain length $\mc L\to \infty$ and  the Matsubara Green's function for spin current operators,
\begin{equation}
{\cal G}_{\alpha\beta}^{ab}\left(x,y;\tau\right)=\left\langle {\cal T}_{\tau}J_{\alpha}^{a}\left(\tau,x\right)J_{\beta}^{b}\left(0,y\right)\right\rangle, \label{spinCurrentGreen}
\end{equation}
where ${\cal T}_\tau$ is the imaginary-time ordering operator. Importantly, this Kubo formula neglects   the resistance at the contacts between the spin chains and the corresponding spin reservoir. For a single ideal chain,  the maximum value of the conductance predicted by this  formula is $G=\frac{1}{4\pi}$, i.e., half of the conductance quantum in (\ref{conductancequantum}). This extra factor 2 can be traced back to the effective Luttinger parameter for the Heisenberg spin chain, which appears in the conductance for a Luttinger liquid only when one neglects the  effects of noninteracting (or Fermi liquid) leads in the dc conductance  \cite{Maslov1995,Meier2003}. We refer the reader to \cite{Oshikawa2006} for a discussion of how the contact to the leads affects the conductance tensor of the electronic Y junction. 

We now address the problem of computing   the correlation function in (\ref{spinCurrentGreen}) in the presence of a conformal boundary condition. Expanding the current operator in terms of the chiral currents, we obtain
\bea
{\cal G}_{\alpha\beta}^{ab}\left(x,y;\tau\right)& = &v^2\left[\left\langle J_{\alpha,L}^{a} (z_1)J_{\beta,L}^{b} (z_2)\right\rangle+\left\langle J_{\alpha,R}^{a} (\bar z_1)J_{\beta,R}^{b} (\bar z_2)\right\rangle\right.\nonumber\\
&&\left.-\left\langle J_{\alpha,L}^{a} (z_1)J_{\beta,R}^{b} (\bar z_2)\right\rangle-\left\langle J_{\alpha,R}^{a} (  z_1)J_{\beta,L}^{b} (\bar z_2)\right\rangle\right],
\eea
where $z_1=v\tau+ix$, $z_2=iy$ and we omit the time ordering operator on the right hand side. 
The absence of  energy and momentum flow across the boundary implies that the two chiral sectors of the bulk CFT are not  independent   \cite{Cardy1989}. Correlation functions between spin currents of the same chirality ($L/R$) retain the bulk form,
\begin{equation}\label{eq:bulkLL}
\left\langle J^a_{\alpha,L} \left(z_1\right)J^b_{\beta,L} \left( z_2 \right) \right\rangle = \frac{1}{8\pi^2} \frac{\delta_{ab}\delta_{\alpha\beta}}{\left(z_1-z_2\right)^2}.
\end{equation}
Conversely, correlation functions for opposite chirality acquire a normalization which depends on the boundary condition $\mc B$ \cite{Cardy1991,Wong1994},
\begin{equation}\label{eq:bdryLR}
\left\langle J^a_{\alpha,R} \left(\bar z_1\right) J^b_{\beta,L} \left( {z}_2 \right) \right\rangle = \frac{\delta_{ab}}{8\pi^2} \frac{A^{\mc B}_{\alpha\beta}}{\left(\bar z_1- {z}_2\right)^2}.
\end{equation}
The coefficients $A^{\mc B}_{\alpha\beta}$ are determined by the  boundary state $\left| \mc B \right\rangle$ associated with the boundary conditions \cite{Cardy1989,Rahmani2010}. Before proceeding with the calculation, we note that equations (\ref{eq:bulkLL}) and (\ref{eq:bdryLR}) imply that for two different chains, $\alpha\ne\beta$, the correlation function for the spin current operator  reduces to
\begin{equation}
	{\cal G}_{\alpha\beta}^{ab}\left(x,y;\tau\right)
	=-\frac{\delta_{ab}}{8\pi^2}\left[\frac{A_{\alpha\beta}^{\mc B}}{\left(\bar{z}_1-z_2\right)^2}+\frac{A_{\beta\alpha}^{\mc B}}{\left(z_1-\bar{z}_2\right)^2}
	 \right]\qquad\qquad (\alpha\neq\beta).
	\label{eq:JalphaJbetacf}
\end{equation}
Inserting the above expression into the Kubo formula (\ref{eq:Kubo}) and performing the integrals, we obtain  
\begin{eqnarray}
	{G}^{ab}_{\alpha\beta}= -\frac{\delta_{ab}A_{\alpha\beta}^{\mc B} }{4\pi}  \qquad\qquad(\alpha\ne\beta).
	\label{eq:Grahmani}
\end{eqnarray}
Thus, the off-diagonal components  of the  conductance tensor  are   determined solely by  the coefficient $ A^{\mc B}_{\alpha\beta}$ in the correlation function  (\ref{eq:bdryLR}) \cite{Rahmani2010,Rahmani2012}.

For the calculation of the conductance tensor, we  first expand the chiral currents  in terms of the operators in the embedding  SU$(2)_3\times\mathbb{Z}_3^{(5)}$.
Inverting (\ref{eq:Currents-embedding-rotation}), we obtain
\begin{equation}
\left(\begin{array}{c}
J_{1,L}^{a}\\
J_{2,L}^{a}\\
J_{3,L}^{a}
\end{array}\right)=\frac{1}{3}\left(\begin{array}{ccc}
1 & \omega & \omega^{*}\\
1 & \omega^{*} & \omega\\
1 & 1 & 1
\end{array}\right)\left(\begin{array}{c}
\mathcal{J}^{a}\\
\phi^{a}_{1}\Psi\\
\phi^{a}_{1}\Psi^{*}
\end{array}\right).\label{eq:Operators-embedding}
\end{equation}
For the right-moving part, we write the same relation but   treat the   operators as the analytic continuation or mirror image of the left-moving part: $\mc O_R(\bar z)=\mc O_R(\tau,x)\mapsto \mc O_L(\tau,-x)=\mc O_L(\bar z)$. 
We can simplify the   expression for the correlators by using the fusion rules and noting that the only terms that give nonzero contributions are those in which the operators can fuse to the identity. For currents in the same chiral sector, we get  
\bea
\left\langle J_{\alpha,L}^{a}\left(z_{1}\right)J_{\beta,L}^{b}\left(z_{2}\right)\right\rangle  &=& \frac{1}{9}\left\langle \mathcal{J}^{a}(z_1)\mathcal{J}^{b}(z_2)\right\rangle  +\frac{1}{9}\left\langle \phi^{a}_{1}(z_1)\phi^{b}_{1}(z_2) \right\rangle\times\nonumber
\\
 &  &\times \left[  \omega^{\alpha-\beta}\left\langle \Psi (z_1)\Psi^{*}(z_2)\right\rangle +\omega^{\beta-\alpha}\left\langle \Psi^{*}(z_1)\Psi(z_2)\right\rangle  \right].\nonumber\\
 \label{eq:Honest2ptF} 
\eea
The normalization of the fields $\boldsymbol \phi_{1}$, $\Psi$ and $\Psi^*$ is fixed so as to recover the correlations of chiral currents for the simplest case of open boundary conditions. Indeed, we then find the free
form anticipated in equation (\ref{eq:bulkLL}) \cite{di1997conformal,Gogolin2004},
\begin{eqnarray}
\left\langle J_{\alpha,L}^{a}\left(z_{1}\right)J_{\beta,L}^{b}\left(z_{2}\right)\right\rangle  & = & \frac{\delta_{ab}}{3}\frac{1+2\cos\left[\frac{2\pi\left(\alpha-\beta\right)}{3}\right]}{8\pi^{2}\left(z_{1}-z_{2}\right)^{2}}
=\frac{\delta_{ab}\delta_{\alpha\beta}}{8\pi^{2}\left(z_{1}-z_{2}\right)^{2}}.
\label{eq:JLJL}
\end{eqnarray}

The mixed left-right correlator depends on the boundary state $|{\mc B}\rangle$. In our case, all the boundary states can be labeled as $|{\mc B}\rangle=|s,f\rangle$, where $s$ labels the primary in the SU(2)$_3$ sector and $f$ the primary in the $\mathbb Z_3^{(5)}$ sector generating the boundary state. In this notation, the O fixed point is identified with the identity, $|\textrm{O}\rangle=|0,\mathbb I\rangle$, while the other fixed points are given by $|\textrm{K}\rangle=|\frac12,\mathbb I\rangle$, $|\textrm{C}_+\rangle=|0,\xi\rangle$, and $|\textrm{C}_-\rangle=|0,\xi^*\rangle$. As for (\ref{eq:Honest2ptF}), the $LR$ correlator can be written as
\begin{eqnarray}
\left\langle J_{\alpha,L}^{a}\left(z_{1}\right)J_{\beta,R}^{b}\left(\bar{z}_{2}\right)\right\rangle  & = & 
\frac{\delta_{ab}}{24\pi^{2}\left(z_{1}-\bar{z}_{2}\right)^{2}}
\left[1+2\textrm{Re}\left(\mathcal{F}^{{\mc B}}\omega^{\alpha-\beta}\right)\right],
\label{eq:JLJR}
\end{eqnarray}
where    $\mc F^{\mc B}$ is defined in analogy to  (\ref{eq:Honest2ptF}) for the correlation functions of the spin-1 primary and the $\Psi$ and $\Psi^*$ fields. Given that the correlation functions of products of fields in different sectors factorize, it can be computed as \cite{Cardy1989,Cardy1991} 
\be
\mathcal{F}^{{\mc B}}=\mathcal{F}^{s,f}=X^{s} Y^f,
\ee
where  $X^{s}$ and $Y^{f}$ are given in terms of the modular $\mathcal{S}$-matrices,
\bea
X^{s}&=& \frac{\mathcal{S}_{s,1}^{ \textrm{\tiny SU(2)}_{3}}\mathcal{S}_{0,0}^{ \textrm{\tiny SU(2)}_{3}}}{\mathcal{S}_{0,1}^{ \textrm{\tiny SU(2)}_{3}}\mathcal{S}_{s,0}^{ \textrm{\tiny SU(2)}_{3}}}\label{LRnormalizationReal},\qquad \qquad
Y^{f}=\frac{\mathcal{S}_{f,\Psi}^{\mathbb{Z}_{3}}\mathcal{S}_{\mathbb{I},\mathbb{I}}^{\mathbb{Z}_{3}}}{\mathcal{S}_{\mathbb{I},\Psi}^{\mathbb{Z}_{3}}\mathcal{S}_{f,\mathbb{I}}^{\mathbb{Z}_{3}}}  .
\label{LRnormalizationComplex}
\eea
The application of the above considerations to the multichannel Kondo problem was given in \cite{Ludwig1994}.

We can now calculate the spin conductance using the Kubo formula (\ref{eq:Kubo}). We  simplify the result using Re$(\omega^{\alpha-\beta})=\frac12(3\delta_{\alpha\beta}-1)$ and Im$(\omega^{\alpha-\beta})=-\frac{\sqrt3}2\varepsilon_{\alpha\beta}$. We obtain $ {G}_{\alpha\beta}^{ab}= {G}_{\alpha\beta}\delta_{ab}$ with
\begin{eqnarray}
 {G}_{\alpha\beta}  & = & \frac{1}{4\pi }
\left\{\left[1-\textrm{Re}\left(\mc F^{\mc B}\right)\right]\left(\delta_{\alpha\beta}-\frac{1}{3}\right)
-\frac{\textrm{Im}\left(\mc F^{\mc B}\right)}{\sqrt{3}}\varepsilon_{\alpha\beta}\right\},
\label{eq:Conductance-Z3chargedBC}
\end{eqnarray}
in agreement with the general expression (\ref{Z3conductance}). For open boundary conditions, $\mc F^{O}=X^0Y^{\mathbb I}=1$
and the conductance vanishes. For the three-channel Kondo fixed point,
we have instead $\mc F^{K}=X^{1/2}=-[4\cos^2(\pi/5)]^{-1}$.
Since $\mc F^{K}\in\mathbb R$, the conductance tensor at the K fixed point is symmetric, \bea
 {G}_{\alpha\beta}  & = & \frac{1}{\pi }
\sin^2\left(\frac{\pi}{5}\right) \left(\delta_{\alpha\beta}-\frac{1}{3}\right)\qquad \qquad (\textrm{K fixed point}).\label{eq:conductance-Kondo}
\eea
Finally, the chiral fixed points must have complex $\mc F^{C_{\pm}}$ in order for  the conductance to have a nonzero antisymmetric part. Since $X^s\in \mathbb R$ $\forall s$, the factor    $Y^f$ must be complex. This is only possible for fusion with $\mathbb Z_3$-charged fields in the $\mathbb Z_3^{(5)}$ sector (see \ref{app:Z3}). In particular, the conjugate fields $\xi$ and $\xi^*$ have charges $+1$ and $-1$, respectively. Using the modular $\mc S$-matrix for the $\mathbb Z_3^{(5)}$ theory,  we find  $\mc F^{C_+}=Y^\xi=\omega^*$ and $\mc F^{C_-}=Y^{\xi^*}=\omega$. More generally, all charged operators will generate an asymmetric conductance, which provides an intriguing physical intuition of this quantum number in the context of Y junctions.
The conductance for the chiral fixed points thus becomes 
\begin{eqnarray}
 {G}_{\alpha\beta}  & = &
 \frac{1}{4\pi } \left[\frac12\left(3\delta_{\alpha\beta}-1\right)\pm\frac12 \varepsilon_{\alpha\beta}\right]\nonumber\\
 &=& \frac{1}{4\pi } \left(\delta_{\alpha\beta}-\delta_{\alpha,\beta\pm1}\right)\qquad \qquad (\textrm{C}_\pm\textrm{ fixed point}).
\label{eq:Conductance-Chiral} 
\end{eqnarray}
This result differs from equation (\ref{conductancehydro}) by a factor 2 because it represents the Kubo conductance calculated without taking into account the contacts to spin reservoirs. The maximally asymmetric conductance tensor (\ref{eq:Conductance-Chiral}) is a direct consequence of the chiral boundary conditions (\ref{bcchiralJ}).   We note in passing that for Y junctions of electronic quantum wires \cite{Oshikawa2006} the maximally asymmetric charge conductance is obtained  only asymptotically for Luttinger parameter $K\to1^{+}$, i.e., for infinitesimal attractive interactions at the edge of stability of the corresponding chiral fixed point.

\subsection{Boundary susceptibility and scalar spin chirality at finite temperature\label{subsec:Temperature-dependence}}

For comparison with the numerical QMC results  in section \ref{subsec:QMC}, in this section we present analytical predictions for the temperature dependence of boundary observables.  We start with the \emph{local boundary susceptibility} which  describes the response to a boundary  magnetic field,
\be
H'=-h S_B^z=-h \sum_{\alpha}S^z_{j,\alpha}. 
\ee
The linear response has the form $\langle S_B^z\rangle=\chi_{\textrm{\tiny loc}}h$, where the local susceptibility $\chi_{\textrm{\tiny loc}}$ is determined by the correlation function
\be
\chi_{\textrm{\tiny loc}}(T) =\int_0^\beta d\tau\, \langle S_B^z(\tau)S_B^z(0)\rangle.\label{chilocal}
\ee
 At the O fixed point, we take $S_B^z={\mc K} \sum_{\alpha}J^z_{\alpha,L}(0)$, where $\mc K$ is a nonuniversal prefactor.  Using the finite-temperature correlation functions \cite{Gogolin2004} at inverse temperature $\beta=1/T$, and  introducing a short-time cutoff $\tau_0$ of order $(J_1)^{-1}$, we obtain 
 \bea\label{ChilocOpen}
\chi^{(\rm O)}_{\textrm{\tiny loc}}&=&\mc K^2\sum_{\alpha,\alpha'}\int_{\tau_0}^{\beta-\tau_0} d\tau\,\left\langle J^z_{\alpha,L}(z=v\tau)J^z_{\alpha',L}(0)\right\rangle\\\nonumber
&=&\frac{3\mc K^2}{8\pi ^2v^2}\int_{\tau_0}^{\beta-\tau_0} d\tau\,\left[\frac{\pi/\beta}{\sin(\pi\tau/\beta)}\right]^2
\simeq\frac{3\mc K^{2}}{4\pi^{2}v^{2}\tau_{0}}\left(1-\frac{\pi^{2}\tau_{0}^{2}}{3}T^{2}+\cdots\right).
\eea
Therefore, at weak coupling $|J_\chi|\ll J_\chi^c$, we expect the local susceptibility to approach a nonuniversal value at zero temperature and  to decrease with a quadratic dependence upon increasing $T$. 
At the chiral fixed point, the dominant contribution comes from the
staggered part of the spin operator, represented by  $S^z_B\sim \sum_\alpha \textrm{Tr}[\tilde g_\alpha(0)\sigma^z]\sim \sum_\alpha\sin[\sqrt\pi\varphi_{\alpha,L}(0)-\sqrt\pi \varphi_{\alpha+1,L}(0)]$. The calculation of the boundary susceptibility in this case involves the two-point function for a   dimension-$1/2$ boundary operator and gives 
\begin{equation}
\chi^{(\rm C)}_{\textrm{\tiny loc}}\sim -\ln (\tau_0T).\label{chilocchiral}
\end{equation}
Therefore, the susceptibility  diverges logarithmically as $T\to0$ at the chiral fixed points.  
Near the K fixed point, the leading operator representing the  boundary spin is   the spin-1 primary field of the SU(2)$_3$ WZW model, $\mathbf{S}_{B}\propto   \boldsymbol{\phi}_1$, with dimension $2/5$  \cite{Ludwig1994}. As a result, the boundary susceptibility diverges as a power law at low temperatures, 
\begin{equation}
\chi^{(\rm K)}_{\textrm{\tiny loc}}\sim  T^{-\frac{1}{5}}.\label{chilocKondo}
\end{equation}
While this is a stronger divergence than at the chiral fixed point, we may anticipate difficulties in cleanly distinguishing (\ref{chilocchiral}) and (\ref{chilocKondo}) from stochastic QMC data.

We now turn to  the thermal average of the \emph{boundary SSCO},
\be
C_B(T)\equiv\left\langle \hat C_{j=1}\right\rangle=Z^{-1}\textrm{Tr}\left(\hat C_{1}e^{-\beta H}\right).
\ee
Clearly, $C_B(T)$ vanishes at any temperature  for $J_\chi=0$ due to time-reversal symmetry. In the strong-coupling limit, $J_\chi\to \pm \infty$, we have $C_B\to \mp \frac{\sqrt3}{4}$ as the chirality saturates at  the eigenvalue of $\hat C_1$. 
In general, from the Hamiltonian (\ref{Htotal}), we see that $C_B$ can be obtained as
\be
C_B(T)=-\frac1\beta\frac{\partial}{\partial J_\chi}\ln Z.\label{CfromlnZ}
\ee
 $J_\chi$ can thus be regarded as  external parameter that couples to the boundary SSCO. In analogy with the response to an external field, we  also define the chirality susceptibility,
\be
-\frac{d C_B}{dJ_\chi}=\int_0^\beta d\tau\, \left\langle \hat C_1(\tau)\hat C_1(0)\right\rangle.
\ee
Near the O fixed point,  perturbation theory in the boundary Hamiltonian (\ref{HBO}) yields the partition function
\be
Z\simeq 1+\frac{3(\lambda_3^{(O)})^2}{(8\pi^2v^2)^3}\beta \int_0^\beta d\tau\, \left[\frac{\pi/\beta}{\sin(\pi \tau/\beta)}\right]^6
,
\ee
where $\lambda_3^{(O)}\propto J_\chi$. Expanding $\ln Z$ for small $J_\chi$ and using (\ref{CfromlnZ}), we find
\be \label{cBO}
C^{(\rm O)}_B(T)\approx C^{(\rm O)}_B(0)\left(1+\frac{5\pi^2\tau_0^2}{3}T^2+\dots\right),
\ee
where $C^{(\rm O)}_B(0)\propto - J_\chi$ is the nonuniversal $T=0$ value. Note that $C^{(\rm O)}_B(T)$ increases quadratically with increasing $T$. Similar contributions to the nonuniversal prefactor of the $T^2$ term may come from the   time-reversal-invariant irrelevant  boundary operators in  (\ref{HBO}), assuming the corresponding  coupling constants are even functions of $J_\chi$. Next, we discuss
the behavior near the chiral fixed point, where $|J_\chi|=J_\chi^c\sim J_1$. We then cannot calculate the chirality by perturbation theory in $J_\chi$ anymore. Nonetheless, we expect that  the leading temperature dependence stems from the relevant perturbation in  (\ref{HBchiral}), with coupling constant $\lambda_1^{(C)}\approx -b (J_\chi-J_\chi^c)$ for $J_\chi\approx J_\chi^c$. Computing the correction to the partition function to second order in $\lambda_1^{(C)}$, we find 
\be \label{cBC}
  C^{(\rm C)}_B(T)\sim  -6b^2(J_\chi-J_\chi^c) \ln(\tau_0 T).
\ee
Therefore, we predict a logarithmic temperature dependence with a sign change of the prefactor around the critical point. We note that the effective coupling constant gets renormalized at low temperatures, $\lambda_{1,\textrm{\tiny eff}}^{(C)}(T)\sim \lambda_1^{(C)}(\tau_0T)^{-1/2}$. Therefore, the perturbative result assuming chiral boundary conditions  is only valid for $T$ in the range
\be
 \tau_0T^*\equiv  (1-J_\chi/J_\chi^c)^2
 < \tau_0 T\ll 1.
 \ee
Here $T^*$ sets the crossover temperature  to the quantum critical regime of this boundary transition. 
Finally, near the K fixed point, we assume that the coupling constants of irrelevant boundary operators in (\ref{HBK}) are  smooth functions of $J_\chi$. The dominant temperature dependence of $C_B(T)$ then stems from the leading irrelevant operator with dimension $7/5$. Applying perturbation theory in $\lambda_1^{(K)}$, we find
\be 
C^{(\rm K)}_B(T)\approx C^{(\rm K)}_B(0)\left(1+c_1T^{9/5}+c_2T^{2}+\cdots\right),
\ee 
where $c_1$ and $c_2$ are nonuniversal constants. The direct calculation from the correlation function gives $c_1>0$ and $c_2<0$. The precise temperature dependence of the boundary chirality in the strong coupling limit depends on the competition between these contributions. 

\section{Numerical results\label{sec:Numerical-checks}}

In order to check our predictions, we employed both DMRG and QMC simulations.
The first technique aims at ground-state properties and is therefore useful
to identify the boundary  fixed points  and the associated scaling
dimensions of operators through the large-distance decay of correlation functions. The second method, instead,  computes equilibrium
expectation values of local  observables at finite temperature, using the effective field theory directly
in the thermodynamic limit.

\subsection{DMRG\label{subsec:DMRG}}

The density matrix renormalization group is one of the most powerful techniques to investigate ground-state
properties of (quasi-)1D quantum lattice systems. The power of 
this method lies on a systematic truncation of the Hilbert space, using the information provided by the reduced density matrix.
Since its original development \cite{White1992}, several DMRG algorithms have been proposed  to study systems 
with different geometries, such as Y junctions \cite{Guo2006,KumarYjunc2016}, finite-width strips \cite{Whitestrip1998,Flavialadders2014}, and two-dimensional systems \cite{White2D2007}. 

\begin{figure}[t]
\centering{
\includegraphics[width=0.44\textwidth]{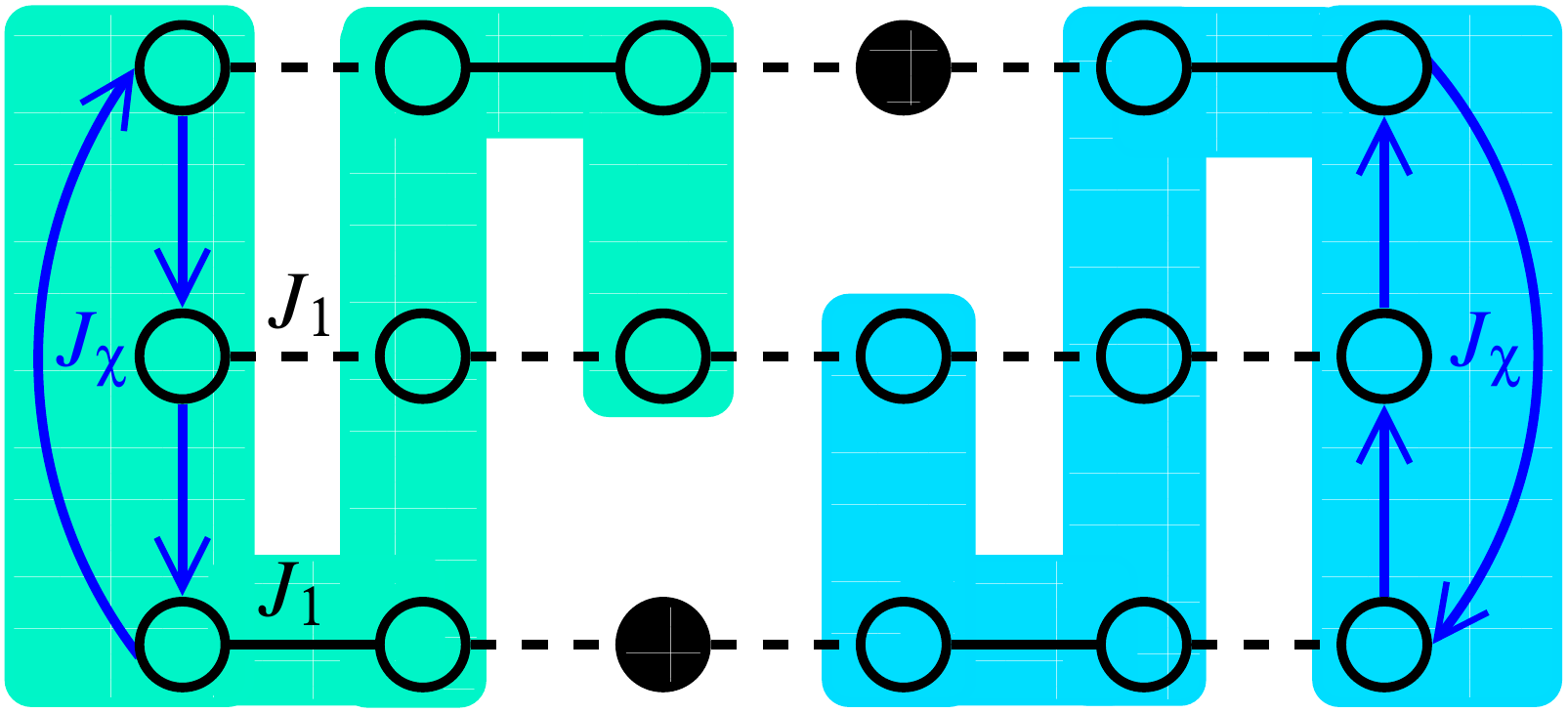}}
\caption{Illustration of the Y junction from the ordinary DMRG point of view. 
The green and blue regions represent the system and environmental DMRG blocks, respectively. The black
solid (dashed) lines represent the nearest-neighbor (long-range) interactions and the blue arrows indicate 
the chiral three-spin interactions with respective ordering at the boundary sites. The filled (open) circles
are the center (renormalized) sites.}
\label{fig4}
\end{figure}

Some results for the  chiral Y junction of spin chains obtained  using the algorithm proposed in \cite{Guo2006}  have already been presented in \cite{Buccheri2018}. The method of  \cite{Guo2006} works efficiently for a Y junction with boundary interaction among spins at the first site, $j=1$ and open boundary conditions at $j=\mathcal{L}$. Here, we apply a different numerical scheme especially tailored for the calculation of the spin conductance, described in \cite{Rahmani2010,Rahmani2012}. In this method, we implement the geometry illustrated in figure \ref{fig3}, featuring two Y junctions facing each other,  where the system size (the length of each chain) is finite and one junction is a mirror image of the other. Implementing this geometry with the method used in \cite{Buccheri2018} is equivalent to considering periodic boundary conditions and would  require a much higher computational effort. Instead, we treat the double Y junction by mapping the system  to a chain with long-range interactions and employing   ordinary DMRG as shown schematically in figure \ref{fig4}. In our DMRG  computations, we have kept up to $m=3000$   states per block. 
 The largest truncation error of our   results at the final sweep is of order $10^{-7}$. 
 In order to check the accuracy  of our correlations under truncation of the Hilbert space, for a fixed system size,
 we compared the numerical data obtained by keeping $m=2400$ versus  $m=3000$ states. The errors
in correlation functions are at least one order of magnitude smaller than the values acquired by DMRG.
 In addition, in our estimates, we did not find   significant differences when fitting the correlations 
 for these distinct numbers of kept states. 
  
In order to characterize the three different regimes of the Y junction  using DMRG, we have calculated  the expectation value of the SSCO at the boundary, the spin  conductance from the Kubo formula, and   the exponent in the power-law decay of the three-spin correlation. In all the DMRG results presented in this subsection, we have set $J_1=1$ and $J_2=0$. Thus, one should keep in mind that the results may be affected by logarithmic corrections due to the marginal operator in (\ref{continuumlimitchain}). Nonetheless, as we discuss below, we find remarkably good agreement with the analytical predictions that neglect logarithmic corrections as well as  with our previous DMRG results \cite{Buccheri2018} for  three-spin correlations in the model with $J_2=J_2^c$.  
 
\begin{figure}[t]
\centering{}
\includegraphics[width=0.70\textwidth]{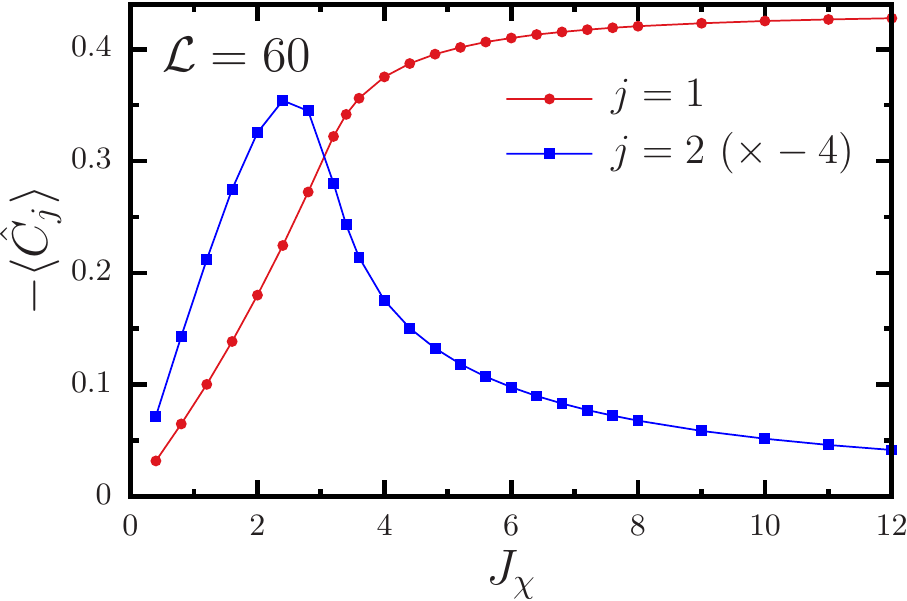}
\caption{DMRG results for the SSCO expectation value, $\bra\hat C_j\ket$, vs $J_\chi$ for $\mc{L}=60$ and two values of $j$. The data for $j=2$ were multiplied  by a factor $-4$.}
\label{fig5}
\end{figure}
   
In figure \ref{fig5}, we show the expectation value
 of the SSCO  at positions $j=1$ and $j=2$  as a function of $J_\chi$ for a Y junction with length $\mathcal{L}=60$. As expected, the chirality at the boundary site $\bra \hat C_1 \ket$ is negative for $J_\chi>0$. Moreover,  its absolute value increases monotonically and approaches  the saturation value 
 $|\langle \hat C_1 \rangle|=\sqrt{3}/4 \approx 0.433$ for $J_\chi\to+\infty$, see equation (\ref{SCenergies}). In contrast, the chirality at the second site, $\bra \hat C_2 \ket$, is positive and reaches a maximum value at intermediate coupling. The saturation of $\bra \hat C_1 \ket$ and the vanishing of $\bra \hat C_2 \ket$   support our picture that   the boundary  spins  form a low-energy  spin-1/2 doublet   and time-reversal symmetry is effectively restored for the remaining spins in the limit   $J_\chi\to\infty$.  The peak in $\bra \hat C_2 \ket$ also provides a rough estimate for the crossover scale separating the weak and strong coupling limits. Around this scale, one expects to find the chiral fixed point. 
 
To pinpoint the location of the chiral fixed  point, we   investigate    the linear-response spin conductance of the Y junction. Here we follow the method developed by 
  Rahmani \emph{et al.}   \cite{Rahmani2010,Rahmani2012}. 
   Although the Kubo formula (\ref{eq:Kubo}) involves a  dynamical correlation function, the dc conductance is uniquely determined by the prefactor of the correlator between $L$ and $R$ currents in (\ref{eq:bdryLR}). We can then rely on conformal invariance to    extract this prefactor     from the large-distance decay of \emph{static} correlation functions, which are easy to access via time-independent DMRG. We then set $\bar z_1=-z_2=-ij$, with $j$ the distance from the boundary, and use the conformal map (\ref{conformalmap}) to account for the  finite system size. Using equation (\ref{eq:Grahmani}) to express the coefficient in terms of the conductance, we can write the $RL$ correlation in the form \begin{equation}
\langle J^z_{\alpha,R}\left(j\right)J^z_{\beta,L}\left(j\right) \rangle= \frac{ G_{\alpha\beta}/G_0}{\left[ 4\mathcal{L} \sin\left(\frac{\pi j 	}{\mathcal{L}} \right)\right]^2}\label{eq:spincond}\qquad\qquad (\alpha\neq\beta).
\end{equation}
From  
equation (\ref{eq:spincond}), we see that the problem of estimating the conductance resides in the computation of the correlation function of chiral currents in the ground state. In order to use DMRG, we need to write the chiral currents 
in terms of the spin operators in the lattice model. We can use the relation to  the magnetization and the spin current 
  in   (\ref{magnop}) and (\ref{spincurrop}) and write 
\begin{equation}
 J^z_{\alpha,R}(j)=\frac{1}{2v}\left[v M^z_{\alpha}(j)+J^z_\alpha(j)\right],
\end{equation}
\begin{equation}
 J^z_{\alpha,L}(j)=\frac{1}{2v}\left[v M^z_{\alpha}(j)-J^z_\alpha(j)\right],
\end{equation}
where $v=\pi /2$ is the spin velocity for the  Heisenberg chains. The magnetization and spin current 
  operators are related to the   spin operators by
\bea
M^z_{\alpha}(j)=\frac{1}{2}\left( S^z_{\alpha, j}+S^z_{\alpha, j+1} \right),\label{magnetizationlattice}\\
J^z_{\alpha}(j)=\frac{i}{2}\left( S^+_{\alpha, j}S^-_{\alpha, j+1}-S^-_{\alpha, j}S^+_{\alpha, j+1} \right).
\eea
The spin operators obey the discrete-space version of the continuity equation in the bulk, $\partial_tS_j^z(t)+J^z_{\alpha}(j)-J^z_{\alpha}(j-1)=0$. The linear combination in (\ref{magnetizationlattice}) is important to cancel out the staggered magnetization to leading order in the mode expansion (\ref{nabSpin}). 

\begin{figure}[t]
\centering
\includegraphics[width=0.47\textwidth]{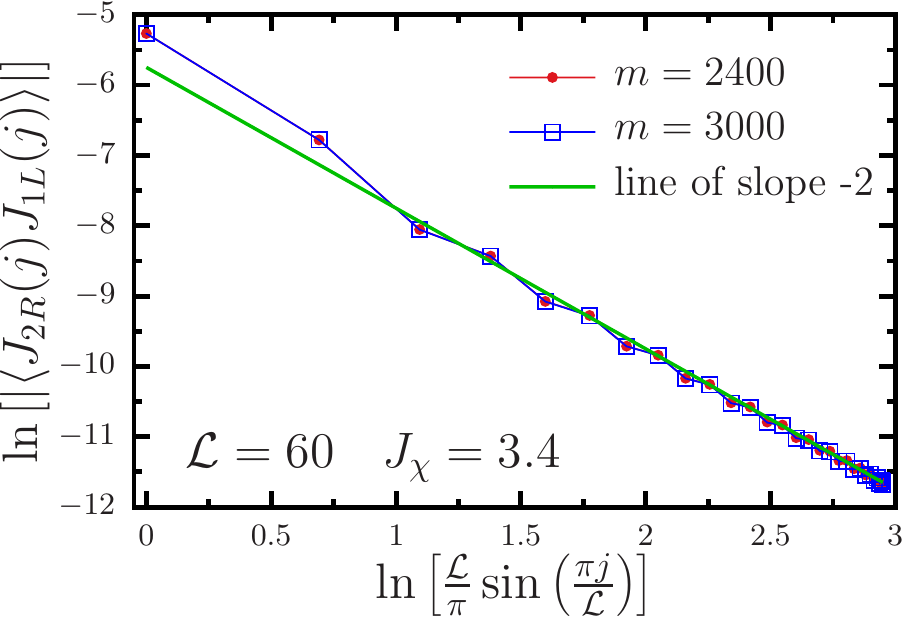} 
\includegraphics[width=0.47\textwidth]{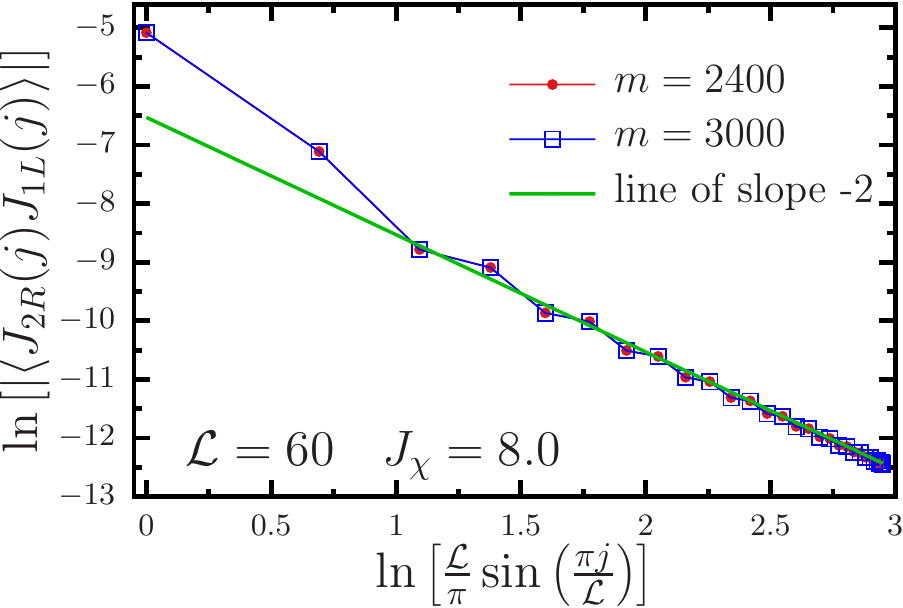}
\caption{Logarithm of the $RL$ correlations vs  
$\mathrm{ln}\left[\frac{\mathcal{L}}{\pi}\sin\left(\frac{\pi j}{\mathcal L}\right)\right]$ for $J_\chi=3.4$ 
(left panel) and $J_\chi=8$ (right panel). In order to show the accuracy of the correlations obtained by
the DMRG, numerical data are shown for two numbers $m$ of kept states.
\label{fig6}}
\end{figure}

The spin conductance is estimated by fitting the 
correlations $\left\langle J^z_{\alpha,R}\left(x\right)J^z_{\beta,L}\left(x\right) \right\rangle$ 
using   equation (\ref{eq:spincond}). Note that   equation (\ref{eq:spincond}) is useful only 
when the  spin conductance is finite. In the case of vanishing $G_{\alpha\beta}$, the current-current correlations do not scale  
linearly with $\left[4\mathcal{L}\sin\left(\frac{\pi j}{\mathcal{L}}\right)\right]^{-2}$. Instead, they are
dominated by contributions of irrelevant operators which are responsible for a faster decay.
It is worth mentioning that  $G_{\alpha\beta}(J_\chi)=G_{\beta\alpha}(-J_\chi)$, but   in general  we have $G_{\alpha\beta}(J_\chi)\neq G_{\beta\alpha}(J_\chi)$ due to the breaking of reflection and time-reversal symmetries.  In particular,  we expect the spin conductance to be  maximally asymmetric at the chiral fixed point. For $J_\chi>0$ (see figure \ref{fig3}), this means that $G_{12}$ (and equivalent components obtained by cyclic permutations of the leg indices) must reach its maximum value at $J_\chi=J_\chi^c$, while it must vanish at $J_\chi=-J_\chi^c$. 

 Indeed, in figure \ref{fig6} we observe a linear behavior of the $RL$ correlations with 
 $\left[4\mathcal{L}\sin\left(\frac{\pi j}{\mathcal{L}}\right)\right]^{-2}$  for values of $J_\chi$ where we expect a finite spin
 conductance. In the same figure,  we   show   DMRG results    for two different numbers of states included in the calculation, showing the robustness of the estimate to truncation errors. 
Since the BCFT prediction (\ref{eq:spincond}) is only valid for distances far from the 
 boundary, when extracting the spin conductance numerically, we made sure that the fitting interval covers only the region which
 exhibits such scaling behavior.

\begin{figure}[t]
\centering
\includegraphics[width=0.7\textwidth]{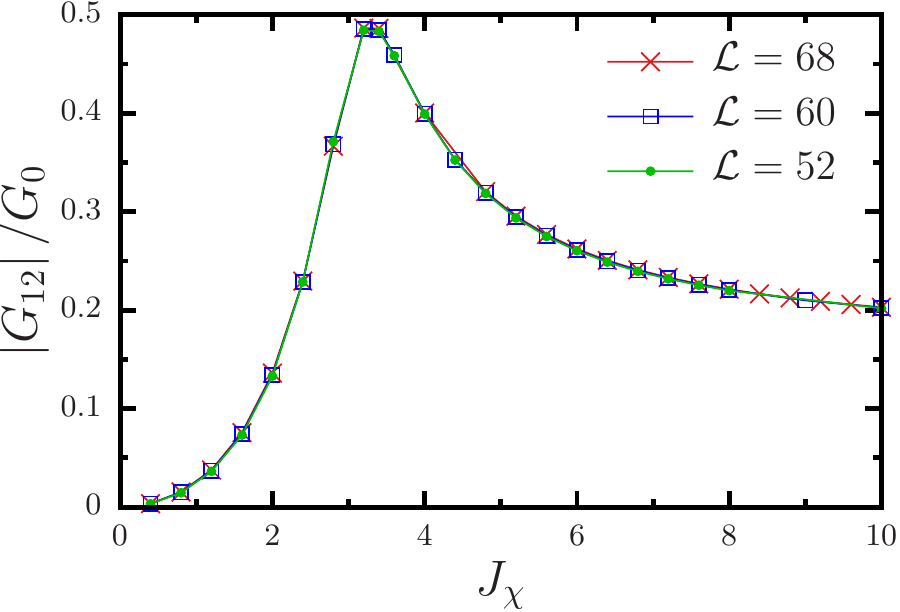}
\caption{DMRG results for the spin conductance $G_{12}$ vs $J_\chi$ for different system sizes. }
\label{fig7}
\end{figure}

In figure \ref{fig7}, we show our estimates of the spin conductance as a function of $J_\chi$ for   different system sizes. The 
predictions for $G_{12}/G_0$ at the O, C$_-$ and K fixed points are 0, $-\frac{1}{2}$ and 
$-\frac{2}{3}\sin^2(\pi/5)$, respectively, see equations (\ref{eq:conductance-Kondo}) and (\ref{eq:Conductance-Chiral}).  
 Note that if we were able to compute the correlation at asymptotically  large distances in the limit ${\mc L}\to \infty$, we would expect the conductance  to be a discontinuous function of $J_\chi$, taking the values $G_{12}=0$ for $J_\chi<J_\chi^c$, $G_{12}/G_0=-1/2$ right at the critical point, and $G_{12}/G_0=-\frac{2}{3}\sin^2(\pi/5)$ for $J_\chi>J_\chi^c$. In contrast, the result for $G_{12}(J_\chi,{\mc L})$ for finite $\mc L$ is a smooth function of $J_\chi$ that must approach the fixed point values as we increase $\mc L$. 

\begin{figure}[t] 
\centering
\includegraphics[width=0.7\textwidth]{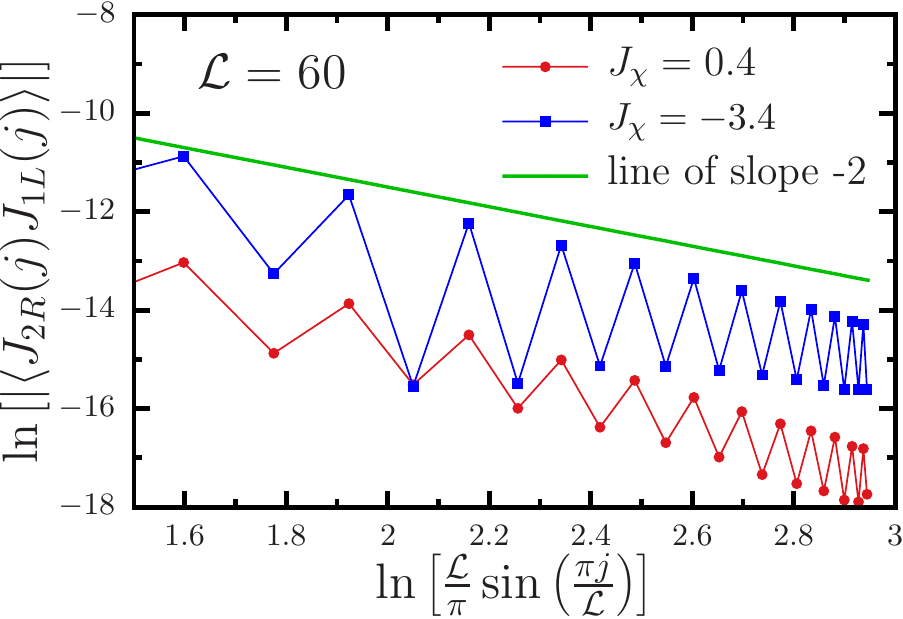}
\caption{Logarithm of the $RL$ correlations obtained by DMRG vs
$\mathrm{ln}\left[\frac{\mathcal{L}}{\pi}\sin\left(\frac{\pi j}{\mathcal{ L}}\right)\right]$, for $J_\chi=0.4$
and $J_\chi=-3.4$. In order to demonstrate that correlations decay faster than  $\left[\sin\left(\frac{\pi j}{\mathcal{L}}\right)\right]^{-2}$, 
we also show a line with slope $-2$.}
\label{fig8}
\end{figure}

From  figure \ref{fig7},  we identify a maximum in the conductance  at $J_\chi\approx 3.4$, where the peak conductance is close to the maximum value, 
$|G_{12}|/G_0= 1/2$, predicted by the Kubo formula. On the other hand, the $RL$ correlation at $J_\chi=-3.4$ decays faster than $1/\sin^2(\pi j/\mathcal{L})$ at large distances, 
see figure \ref{fig8}.  
This is the same behavior as observed for $J_\chi=0.4$, where we expect the conductance to vanish in the limit ${\mc L}\to\infty$ because the regime of  small $J_\chi$ is governed by the O fixed point.  These results indicate that 
for $\mathcal{L}\to \infty$, $G_{12}$ will vanish for $J_\chi=-3.4$, and hence $G_{21}$ vanishes for $J_\chi=+3.4$ as well. 
Altogether this provides strong evidence that the C$_-$ fixed point is located at $J^c_\chi\approx 3.4$.  
This estimate for the nonuniversal critical coupling value differs only slightly from the one reported in our earlier work
($J_\chi^c \approx 3.1$) \cite{Buccheri2018}, where the case $J_2=J_2^c$ has been studied instead of the model with $J_2=0$ considered here.

\begin{table}[t]
  	\centering
 	\caption{Comparison of the (absolute value of the) dimensionless spin conductance $G_{12}/G_0$ for three values of $J_\chi$ corresponding to the O, C and K points, respectively. The DMRG estimates for $J_\chi=0.4$ and $3.4$ were obtained for $\mathcal{L}=68$. The value for $J_\chi=10$ has been extrapolated to infinite size, as explained in the main text.}
 
 	\vspace*{0.5cm}
 	\label{conductable}
 	\begin{tabular}{| c @{\quad} |c @{\quad} |c @{\quad}|c @{\quad} |c |} \hline
 		Fixed Point & $J_\chi$& DMRG & BCFT  & rel. error\\ \hline
 		 O         & $0.4$    & $0.004$  &  $0$    &         \\
 		 C         & $3.4$     & $0.498$   &  $0.5$     &  $0.4\%$      \\
 		 K         & $10$    & $0.217$   &  $0.2303\ldots$   & $5.7\%$    \\ \hline
  	\end{tabular}  
 \end{table} 

 We now turn to the determination of the conductance at the K fixed point. For $\mc L\to \infty$, our expectation is that the off-diagonal conductance should approach the plateau value corresponding to the K fixed point, for $J_\chi>J_\chi^c$. Conversely, for any finite size, the conductance is a continous function of the coupling $J_\chi$, approaching smoothly an asymptotic value.
  We have observed that the fit of the current-current correlation function by the form (\ref{eq:spincond}) is not reliable for $J_\chi>10$, due to a combination of truncation errors, a residual staggered contribution and an evident dependence of the result on the fitting interval. On the other hand, as clear also from figure \ref{fig5}, the effective central spin is already fully developped at $J_\chi=10$, as the expectation value of the SSCO at the boundary has reached over $98\%$ of its asymptotic value for all system sizes; see discussion around (\ref{SCenergies}).
   We therefore select this point as a representative of the strong coupling regime.
 Compared to the O and C points, finite-size effects are noticeably more important for large $J_\chi$. The values of the conductance $G_{12}(J_\chi,\mc{L})$ at $J_\chi=10$ show a slow, but significant variation with the system size $\mc{L}$. In order to extract the infinite-size limit, we considered the values of the conductance for different system sizes $\mathcal{L}=52,\,60,\,68$ and performed an extrapolation in $\mathcal{L}$, assuming the form $G_{12}(J_\chi,\mc{L})=G_{12}(J_\chi,\infty)+a_1 \mc{L}^{-a_2}$, with free fitting parameters $a_1$ and $a_2$. The asymptotic value $G_{12}(10,\infty)$ is quoted in table \ref{conductable}.
 Given the small fitting interval in $\mc L$ and the DMRG truncation errors, this result should be taken with some precaution.  Nonetheless, the available evidence from DMRG is consistent with the three-channel Kondo fixed point scenario.

\begin{figure}[t]
\centering
\includegraphics[width=0.7\textwidth]{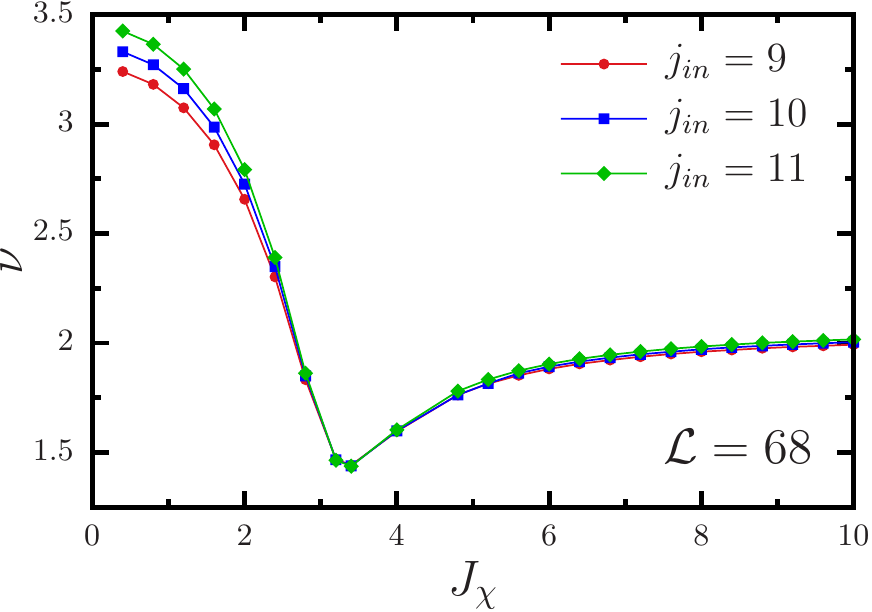}
\caption{Exponent $\nu$ of the power-law decay of the three-spin correlations 
$G_3(j)$ vs $J_\chi$. The shown results were obtained from fits to DMRG data in the
interval $j_{in}\leq j \leq 32$. The minimum is around the chiral fixed point, in 
agreement with the theoretical prediction (\ref{CnnDecayChi}).  }
\label{fig9}
\end{figure}

 Let us next discuss  DMRG results for the three-spin correlation $G_3(j)$ described in section \ref{subsec:Spin-chirality-operator}. This quantity oscillates with the  distance from the boundary and exhibits  power-law decay with an exponent governed by the low-energy fixed point, see equations (\ref{G3oscdecay}) and (\ref{G3exponents}).  Using the conformal transformation (\ref{conformalmap}), the three-spin correlation functions are brought into the form
   \begin{equation}
 G_3(j)\sim \frac{(-1)^j}{ \left[\frac{\mathcal{L}}{\pi}\sin\left(\frac{\pi j}{\mathcal{L}}\right)\right]^\nu} .
 \end{equation}
  We have fitted our numerical results to this formula and thereby extracted the exponent $\nu$. 
 In figure \ref{fig9}, we show our estimates for $\nu$ as a function of 
 $J_\chi$. As discussed for the conductance, the function  $\nu(J_\chi,\mc L)$ varies continuously with $J_\chi$ for finite $\mc L$. In order to fit the data, we chose an interval   $j_{in}<j<\mc L/2$ such that sites which are too close to the boundary or to the center of the double junction are not taken into account.
We observe a robust  minimum in $\nu(J_\chi)$ at $J_\chi\approx 3.4$,  in remarkable agreement with the location of the maximum in the spin conductance. We have   verified that this minimum is insensitive to changes of the values of $j_{in}$ and/or ${\mc L}$.  Moreover, the value of $\nu$ at the minimum is rather close to the prediction  for the chiral fixed point, $\nu_C=1.5$, see equation (\ref{G3exponents}). 
 Finally, for estimating the exponent $\nu_K$ near the K fixed point, we consider again $J_\chi=10$ as a representative, as the limitations noticed for the conductance apply also to the three-spin correlation. Compared to the spin conductance, finite-size effects are here much smaller, well below truncation errors, and the largest system size under study ($\mathcal{L}=68$) already provides an accurate answer. In table \ref{exponentstable}, we collect our results for $\nu$ at the three fixed points, with the respective BCFT predictions and relative errors. Overall, we conclude that these numerical estimates are consistent with our analytical predictions.
 
\begin{table}[t]
  	\centering
 	\caption{ The exponents $\nu$ of the three-spin correlation $G_3(j)$ for representative values of $J_\chi$ corresponding to the three fixed points.  The estimates were obtained from DMRG data for $\mathcal{L}=68$, see main text.  BCFT predictions and relative errors are also shown.   \label{exponentstable}}
 	
 	\vspace*{0.5cm}
 	\begin{tabular}{| c @{\quad} |c @{\quad}| c @{\quad}|c @{\quad} |c|}\hline
 		Fixed Point& $J_\chi$  & $\nu$  & BCFT & rel. error \\ \hline 
 		 O         & $0.4$     & $3.31$ & $3.5$           &  $5.4\%$       \\
 		 C         & $3.4$     & $1.44$ & $1.5$           &  $4\%$       \\
 		 K         & $10$    & $2.00$ & $2.1$           &  $4.7\%$       \\ \hline
  	\end{tabular}
 \end{table}

\subsection{QMC simulations\label{subsec:QMC}}

Next we turn to finite-temperature path-integral Monte Carlo simulations.  Our 
QMC scheme  employs a bosonized functional
integral representation of the partition function, where the Gaussian
bulk boson modes are integrated out analytically and the limit of
infinite chain length can be taken from the outset. 
A similar method has previously been applied to study Kondo effects in Luttinger liquids \cite{Egger1998}, and we here describe a generalization of that approach for our Y junction problem.
We will study  the
local spin susceptibility at the junction, $\chi_{{\rm loc}}(T)$, and the boundary scalar spin chirality, $C_B(T)=\langle\hat{C}_1\rangle$, see section \ref{subsec:Temperature-dependence} for analytical predictions near the different fixed points.  
Importantly, our QMC approach is free from sign problems.

\subsubsection{Simulation scheme.}

We start from the bosonized field theory and express the partition sum as imaginary-time functional integral,
 \begin{equation}
 Z = \int {\cal D}[\theta_\alpha(x,\tau)] e^{-S_0[\theta]-J_\chi\int_0^\beta d\tau C_1[\theta]}.
 \end{equation}
 The full action, $S=S_{0}+S_{B}$, contains a Gaussian bulk term ($S_{0}$)
 and a boundary term $S_{B}$ due to $H_{B}$.  We here put the boundary at $x=a$, with a short-distance cutoff $a$, and impose hard-wall boundary conditions at $x=0$.
After integration over the bulk $(\theta_\alpha,\phi_\alpha)$ modes with $x\ne a$, the action $S_{0}$ is
effectively replaced by a time-nonlocal dissipative action, $S_0\to S_{\rm eff}$, for
the boson fields at the position $x=a$.
We now define complex functions of the bosonic Matsubara frequency $\omega$,
\bea
F_\alpha(\omega) &=& 2\int_0^\infty\frac{dk}{2\pi} \frac{\sin(ka)}{k} \tilde\theta_\alpha(k,\omega), \nonumber\\
G_\alpha(\omega) &=& 2\int_0^\infty\frac{dk}{2\pi} \cos(ka)\tilde\theta_\alpha(k,\omega) , \label{fgdef}
\eea
where $\tilde\theta(k,\omega)$ denotes the Fourier transform of the boson field $\theta_\alpha(x,\tau)$.
The real-valued dual boson fields at $x=a$ follow as
\begin{eqnarray}
\theta_\alpha(a,\tau)&=&T\sum_\omega e^{i\omega \tau} G_\alpha(\omega)=G_\alpha(\tau),\nonumber\\ 
\phi_\alpha(a,\tau)&=&T\sum_\omega (-i\omega) e^{i\omega\tau} 
F_\alpha(\omega)=-\partial_\tau F_\alpha(\tau).
\end{eqnarray}
After some algebra, we obtain the effective action in the form
\begin{equation}
S_{\rm eff} =\frac{1}{2} \int_0^\beta d\tau \int_0^\beta d\tau'\sum_\alpha \left( 
\begin{array}{c} G_\alpha(\tau) \\ F_\alpha(\tau) \end{array} \right)^T
{\bf K}(\tau-\tau')
\left( \begin{array}{c} G_\alpha(\tau') \\ F_\alpha(\tau') \end{array} \right).
\end{equation}
The real-valued kernel, ${\bf K}(\tau)=T\sum_\omega \cos(\omega\tau) \tilde {\bf K}(\omega)$, is the Fourier transform of the non-negative matrix kernel 
\begin{equation}
 \tilde {\bf K}(\omega) = \frac{|\omega|}{a|\omega| \coth(a|\omega|)-1}
\left(\begin{array}{cc} \frac{2a|\omega|}{1-e^{-2a|\omega|}} -1  & -|\omega|
\\  -|\omega|& \omega^2 \coth(a|\omega|) \end{array}\right). 
\end{equation}
Importantly, no sign problem arises. Using a hard energy cutoff $D=v/a$, we keep all 
Matsubara components $F(\omega)$ and $G(\omega)$ with $|\omega|<D$.  
Our QMC code uses the full action,  $S=S_{\rm eff}+S_B$, for the MC sampling.
Each data point reported below has been obtained from $\approx 10^7$ to $10^8$ statistically independent samples.

We now present our QMC results for $C_B(T)$ and $\chi_{\rm loc}(T)$. We use units with $D=v/a=1$. Due to our regularization scheme in the low-energy theory behind the QMC approach, the values quoted for $J_\chi$ in this subsection differ from the corresponding values for the lattice model.   

\subsubsection{QMC results.}

First, for very small $J_\chi\ll 1$, we have established  that the analytical results in equations (\ref{cBO}) and (\ref{ChilocOpen}), which are valid near the O fixed point, are accurately reproduced by our QMC data. In fact, for various different $J_\chi\ll 1$, our data (not shown here)  nicely fit the analytical expressions with the same choice for the product $D\tau_0\simeq 3.5$.  This provides an important benchmark test for our scheme.  

\begin{figure}[t]
\centering
\includegraphics[width=0.48\textwidth]{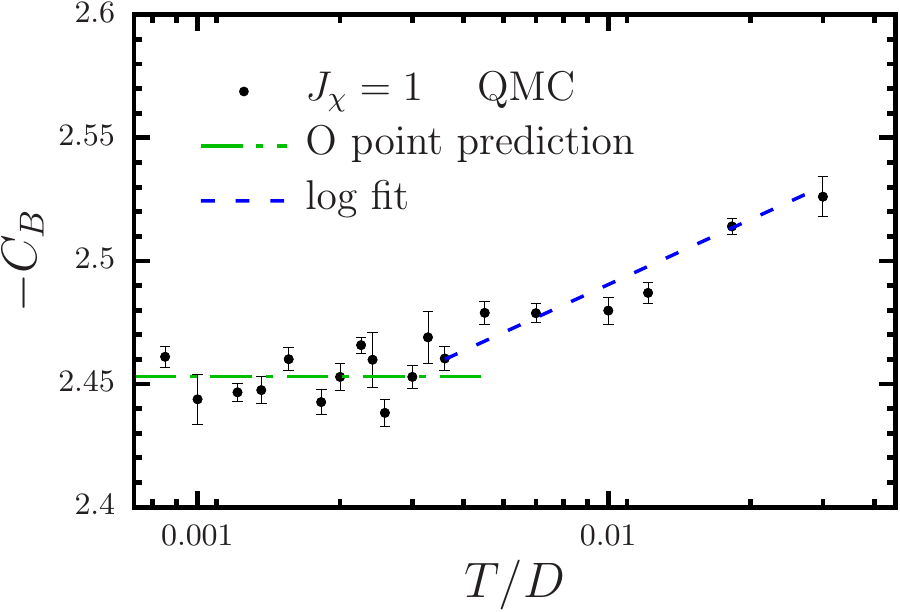}
\includegraphics[width=0.459\textwidth]{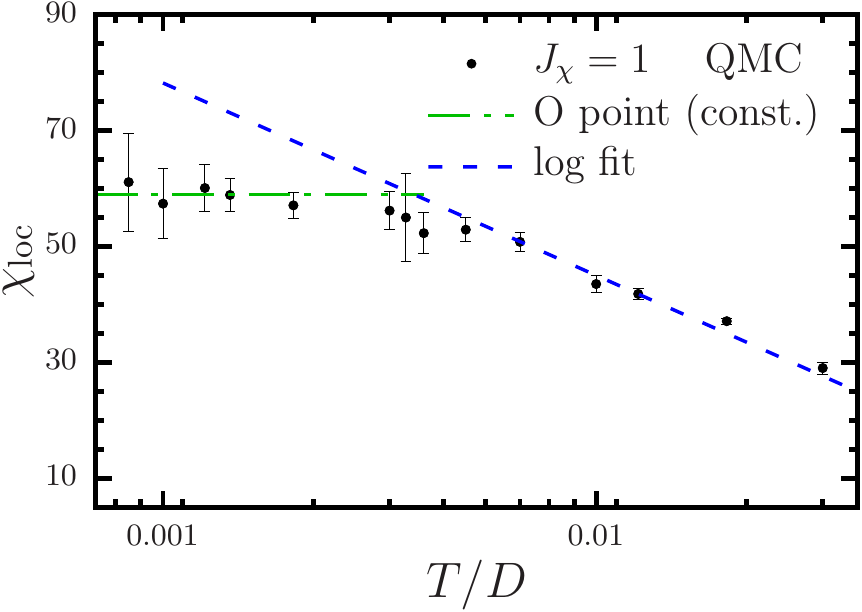}
\caption{QMC data for $J_\chi=1$ (in units with $D=v/a=1$).  Left panel: Scalar spin chirality at the boundary, $C_B$, vs
$T$. Notice the logarithmic $T$ scale. Error bars for QMC data points denote one standard deviation due to the stochastic sampling.  
The green dot-dashed line indicates the expected low-$T$ behavior (\ref{cBO}) near the O fixed point, which essentially predicts a constant-in-$T$ behavior in this temperature regime. The blue dashed curve is a logarithmic fit corresponding to the chiral fixed point, see equation (\ref{cBC}).   
 Right: Local boundary susceptibility,
$\chi_{\rm loc}$, as a function of $T$ for $J_{\chi}=1$. 
\label{fig10}}
\end{figure}

For larger $J_\chi$, we obtain substantial renormalizations of $C_B(T)$ and $\chi_{\rm loc}(T)$ as compared to equations (\ref{cBO}) and (\ref{ChilocOpen}), respectively. In figure~\ref{fig10} we show the corresponding QMC results for $J_\chi=1$.   First, the left panel shows $C_B(T)$ with logarithmic scales of the $T$ axis.  
At elevated temperatures, we observe a logarithmic scaling as
predicted near the chiral fixed point in equation (\ref{cBC}).  This logarithmic scaling thus is interpreted as  high-temperature signature of the unstable chiral fixed point.  The positive slope indicates that $J_\chi=1$ is below $J_\chi^c$, see equation (\ref{cBC}).  At sufficiently low temperatures, the stable O point ultimately dominates and we find a crossover toward the essentially constant $T$-dependence of $C_B$ predicted by 
equation (\ref{cBO}) near the O point. The right panel in figure \ref{fig10} shows $\chi_{\rm loc}(T)$ for the same value of $J_\chi$ (and also with a logarithmic $T$ axis).
For high $T$, we again observe a logarithmic scaling of $\chi_{\rm loc}$ as expected near the chiral fixed point, see equation (\ref{chilocchiral}).
In accordance with the analytical result, here the slope is negative. At lower $T$, the susceptibility saturates to a constant value, in agreement with equation (\ref{ChilocOpen}) valid near the O point. ($T^2$ corrections cannot be resolved within  error bars.)  To summarize, the data in figure \ref{fig10} for $J_\chi=1$ show that although the low-$T$ behavior is dominated by the O point, the high-$T$ behavior is already governed by the unstable chiral fixed point.

\begin{figure}[t]
\centering
\includegraphics[width=0.48\textwidth]{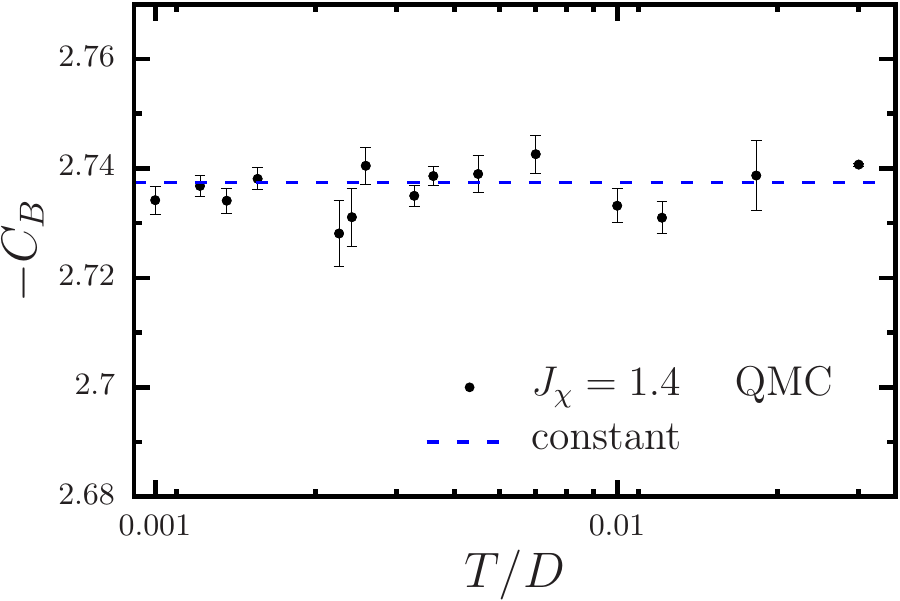} 
\includegraphics[width=0.4675\textwidth]{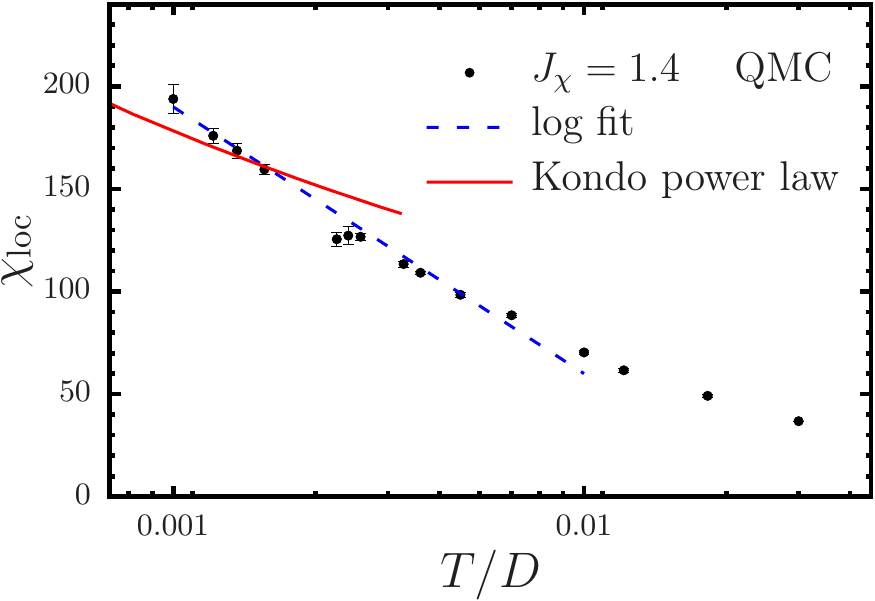}
\caption{QMC data for $J_\chi=1.4$.  Note the logarithmic $T$ scales. Left panel: Scalar spin chirality at the boundary, $C_B$, vs $T$. The blue dashed line is a fit of the  QMC data to a constant. 
Right panel: Local boundary susceptibility,
$\chi_{\rm loc}$, vs $T$. Except at very high $T$, the logarithmic scaling (blue dashed line) expected near the chiral fixed point, see equation (\ref{chilocchiral}), is consistent with the QMC data in the accessible $T$ window. Note that the  power-law scaling near the Kondo fixed point (red solid curve), see equation (\ref{chilocKondo}), is  not consistent with the data. 
 \label{fig11}}
\end{figure}

Next we turn to the value $J_\chi=1.4$, where the corresponding QMC data are shown in figure \ref{fig11}.
This value appears to be quite close to $J_\chi^c$ (in the low-energy theory). Indeed, our data for both $C_B(T)$ and $\chi_{\rm loc}(T)$ are consistent with the respective analytical expressions near the chiral fixed point. The left panel shows that $C_B(T)$ is basically constant, corresponding to a very small prefactor in front of the logarithm in  equation (\ref{cBC}). This is precisely as expected in the vicinity of the chiral point. The right panel shows the local boundary susceptibility, which exhibits a logarithmic $T$ scaling over more than a decade. The power law (\ref{chilocKondo}), which is expected near the Kondo fixed point, is clearly inconsistent with the data.  We conclude that $J_\chi=1.4$ represents a boundary coupling in the near vicinity of the unstable fixed point.
Unfortunately, probing even larger $J_\chi$ by means of QMC simulations turned out to be prohibitively costly. 
We therefore are not able to show results in the Kondo regime using this technique.

\section{Conclusions}\label{sec:Conclusions}

In this paper, we have presented a detailed characterization of the boundary phase diagram of a Y junction of Heisenberg spin chains with a chiral boundary interaction $J_\chi$. Using bosonization  and boundary conformal field theory to construct a low-energy effective field theory,
we  have provided analytical predictions for the boundary entropy,
the spin conductance, and for the low-temperature behavior of the boundary scalar spin chirality  and of the local boundary susceptibility.
The phase diagram exhibits two stable fixed points, namely a fixed point of disconnected chains    at weak coupling (O) and a three-channel Kondo fixed point (K) at strong coupling. These stable points are separated at intermediate
coupling by an unstable chiral fixed point (C).  In BCFT language, the chiral fixed point is described by fusion with  a $\mathbb{Z}_{3}$-charged operator in the $\mathbb{Z}_{3}^{(5)}$ theory associated with  the `flavor' degree of freedom of the junction.

 Using a DMRG scheme  especially suitable for computing the spin conductance,
we have tested the predicted phase diagram. In particular, the chiral fixed point is characterized by a maximally asymmetric spin conductance tensor, which has been unequivocally observed in our numerical calculations, even at small sizes of the system. In comparison, finite-size effects are more pronounced at strong coupling, which causes obstacles to the accurate numerical computation of the spin conductance in this limit. Nonetheless, we find good agreement between the BCFT and the DMRG predictions. 
 In addition, by means of QMC calculations, we have probed the temperature dependence of the local spin susceptibility and the scalar spin chirality at the boundary.  The reported results are in qualitative agreement with our analytical predictions.

The Heisenberg spin chain is known to accurately describe a number of effectively 1D crystalline materials and its excitation spectrum has been probed by many experiments over the years. In view of the rapid developments in antiferromagnetic spintronics and given that spin currents can be readily generated and detected, it stands to reason that our setup can be experimentally realized and investigated  in solid state and/or cold atom platforms. 
Once realized, one would have access to a circulator for spin currents. To the best of our knowledge, spin circulators have not yet been achieved, and hence this would represent a tremendous advancement in the control and manipulation of spin currents.

\appendix

\section{Bosonization}\label{app:WZW}
Here we recall some useful formulas commonly used when studying one-dimensional systems via bosonization \cite{Gogolin2004}.
Let us start with  the case $J_\chi=0$, where we have an open boundary and incoming chiral spin currents are fully reflected,
\be
\mathbf J_{\alpha,R}(x=0)=\mathbf J_{\alpha,L}(x=0)\;.
\ee
One can see that the flow across the boundary of the full spin current 
\begin{equation}
	\mathbf{J}_{\alpha}(x)=v\left[ \mathbf{J}_{\alpha,R}(x) - \mathbf{J}_{\alpha,L}(x)\right]
\end{equation}
is vanishing, with open boundary conditions. It is, in general, convenient to regard $\mathbf J_{\alpha,R}  $ as the analytic continuation of $\mathbf J_{\alpha,L} $ to the negative-$x$ axis,
\be
\mathbf J_{\alpha,R}(x)=\mathbf J_{\alpha,L}(-x),\qquad x\geq0. \label{openBC}
\ee
The effective Hamiltonian  for a given chain  in (\ref{continuumlimitchain}) (with $\gamma=0$) then becomes equivalent to a single chiral mode on the infinite line. The OPE of the right currents can be evaluated using (\ref{KMope}, replacing $z\to\bar z$.

For a description of the low-energy physics, spin operators are bosonized as \cite{Gogolin2004} 
\begin{equation}
\mathbf{S}_{j,\alpha}\sim\mathbf{J}_{\alpha,L}(x=j)+\mathbf{J}_{\alpha,R}(x=j)+\left(-1\right)^{j}\mathbf{n}_\alpha(x=j),\label{nabSpin}
\end{equation}
where the staggered part, 
\begin{equation}
n^{a}_\alpha(x)=\mathcal{A}\mbox{Tr}\left[g_\alpha(x)\sigma^{a}\right],\label{staggered}
\end{equation}
involves the standard Pauli matrices $\sigma^{a}$ and the SU(2) matrix field $g_\alpha(x)$   \cite{di1997conformal}. 
 The SU(2) invariant trace of the matrix field appears in the dimerization operator,
\be
\mathbf S_{j,\alpha} \cdot \mb S_{\alpha,j+1}\sim \textrm{const.}+(-1)^j\mc A'\,\textrm{Tr}[g_\alpha(x)],
\ee
where $\mc A'$ is another nonuniversal constant.


The chiral spin currents, with $J_{\alpha,L}^\pm=J^1_{\alpha,L}\pm i J_{\alpha,L}^2$, are given in terms of the chiral bosonic fields satisfying (\ref{chiralvarphicomm}) by \cite{Eggert1992} 
\begin{equation}
J_{\alpha,L/R}^{\pm}(x)=\frac{1}{2\pi}e^{\pm2i\sqrt{\pi}\varphi_{\alpha,L/R}(x)},\qquad\quad J_{\alpha,L/R}^{z}(x)=\pm\frac{1}{2\sqrt{\pi}}\partial_{x}\varphi_{\alpha,L/R}(x),\label{J2phi}
\end{equation}
while the staggered part takes the form 
\begin{equation}
n_\alpha^{\pm}(x)=\mathcal{A}e^{\pm i\sqrt{\pi}\left(\varphi_{\alpha,L}+\varphi_{\alpha,R}\right)},\qquad\quad n_\alpha^{z}(x)=\mathcal{A}\sin\left[\sqrt{\pi}\left(\varphi_{\alpha,L}-\varphi_{\alpha,R}\right)\right].\label{n2phi}
\end{equation}
The dimerization operator involves
\be
\textrm{Tr}[g_\alpha(x)]= \cos[\sqrt\pi(\varphi_{\alpha,L}-\varphi_{\alpha,R})].\label{dimer}
\ee
 Here $\mc C=-\sqrt\pi$ is equivalent to  $\mc C=\sqrt\pi$ because the boson fields are compactified
 . The  two choices for $\mc C$ correspond to stronger bonds on either even or odd links whenever the dimerization field (\ref{dimer}) acquires a nonzero expectation value.

Within nonabelian bosonization, the currents (\ref{J2phi}) are instead level-1 descendents in the module of the identity \cite{di1997conformal}. Here, we only need to recall that the SU(2)$_k$ theory possesses $k+1$ WZW primary fields labeled by integrable representations of SU(2) \cite{Affleck1991b,di1997conformal}, i.e., by the spin quantum number   $s=0,\frac12,1,\dots, \frac{k}2$. The corresponding scaling dimensions are
\be
\Delta_{s}=\frac{s\left(s+1\right)}{k+2}\label{eq:Dimensionssu(2)_3}.
\ee
The general form of the short-distance expansion of the primary fields
is encoded into the fusion coefficients
$\mathcal{N}_{s_{1},s_{2}}^{s_{3}}$ \cite{Zamolodchikov1986},
\begin{equation}
\mathcal{N}_{s_{1},s_{2}}^{s_{3}}=\left\{
\begin{array}{cl}
1, & \textrm{if }s_{1}+s_{2}+s_{3}\in\mathbb{N} \\
&\textrm{and }\left|s_{1}-s_{2}\right|\le s_{3}\le\min\left(s_{1}+s_{2},\frac{k}{2}\right),\\
0, & \textrm{otherwise.}
\end{array}\right.
\label{su(2)fusionCoefficients}
\end{equation}

\section{$\mathbb{Z}_{3}^{(5)}$ toolbox }\label{app:Z3}

Here we collect some notions about the $\mathbb{Z}_{3}^{(p)}$ CFT, focusing on one chiral sector in particular and $p=5$.
 When $p$ is large, the models $\mathbb{Z}_3^{(p)}$ can be interpreted as critical solutions to the bosonic field theory defined by the action
\begin{equation}
\mathcal{S}=\int d^2 r \left[\partial_\mu \phi^\dagger \partial_\mu \phi + V\left( \phi,\phi^\dagger \right)\right],
\end{equation}
where the polynomial $V$ is invariant under the $\mathbb{Z}_3$ transformation,
\begin{equation}
\phi\to\omega\phi,\qquad \phi^\dagger\to\omega^*\phi^\dagger,\qquad \omega=e^{2\pi i/3}
\end{equation}
has highest degree $p-2$ in $(\phi^\dagger\phi)$, and its parameters have been tuned to a multi-critical point \cite{Fateev1987}.
In addition to the energy-momentum tensor $T(z)$, which generates the conformal
transformations, the theory contains the additional local spin-$3$ currents $W(z)$.
In the same way as the modes $L_n$ of the energy-momentum tensor 
$T(z)=\sum_{n\in\mathbb{Z}} z^{-n-2} L_n$ generate the Virasoro algebra,
the modes $W_n$ of $W(z)=\sum_{n\in\mathbb{Z}} z^{-n-3} W_n$ generate an additional symmetry algebra, denoted by $\cal W$ algebra \cite{Fateev1987}.
The space $\mathcal{A}$ of local fields can be decomposed as
\begin{equation}
\mathcal{A} = \oplus_i \left[\Phi_i\right] ,
\end{equation}
where $\left[\Phi_i\right]$ denotes an irreducible representation of $\cal W$.
In particular, the representation $\left[\Phi_i\right]$ can be constructed starting from an `ancestor' (or primary) field $\Phi_i$,
satisfying the properties
\begin{eqnarray}
L_{n>0} \Phi_i=W_{n>0} \Phi_i =0,\qquad L_0 \Phi_i=\Delta_i \Phi_i ,\qquad W_0 \Phi_i=w_i \Phi_i .
\end{eqnarray}
for some real $w_i$ and non-negative $\Delta_i$.

The theory can be formulated in the Coulomb-gas formalism \cite{di1997conformal},
in terms of a two-component free massless bosonic field $\mathbf{\varphi}=\left(\varphi_1,\varphi_2\right)$.
A background charge makes the $U(1)$ symmetry of the bosonic field theory anomalous and alters the central charge and the scaling dimension of the vertex operators. 
The energy-momentum tensor is written as
\begin{equation}
T=-\frac{1}{4}\sum_{j=1,2}(\partial_z\varphi_j)^2+i\alpha_0\partial_z^2\varphi_1,
\end{equation}
for $\alpha_0=1/\sqrt{30}$. Primary fields $\Phi^{n,m}_{n',m'}$ are labeled by two pairs of integers $n,n'$ and $m,m'$, such that $n+n'\le4$ and $m+m'\le5$. They can be written as free-field vertex operators
\begin{equation}
V_{\mathbf{\beta}}(z)=	V_{(\beta_1,\beta_2)}(z)=e^{i \mathbf{\beta}\cdot\mathbf{\varphi}},
\end{equation}
where $\mathbf{\beta}=\mathbf{\beta}^{n,m}_{n',m'}$ is given in terms of the 
$su(3)$ weights,
\begin{equation}
{\bm\omega}_1=\frac{1}{2}\left(1,\frac{1}{\sqrt{3}}\right),  \qquad\quad
{\bm\omega}_2=\frac{1}{2}\left(1,-\frac{1}{\sqrt{3}}\right),
\end{equation}
and the two real solutions $\alpha_{\pm}$ of the equations
\begin{equation}
\alpha_+\alpha_-=-\frac{1}{4},\qquad\alpha_{+}+\alpha_{-}=\frac{\alpha_0}{2} .
\end{equation}
In particular \cite{Fateev1987}, one has
\begin{eqnarray}\nonumber
\mathbf{\beta}^{n,m}_{n',m'} &=&
2\left[(1-n)\alpha_{+}+(1-m)\alpha_{-}\right]{\bm\omega}_1 \\
&+&\left[(1-n')\alpha_{+}+(1-m')\alpha_{-}\right]{\bm\omega}_2.
\end{eqnarray}
In this formalism, one identifies the fields
\bea
\Phi^{n,m}_{n',m'}=\Phi^{5-n-n',6-m-m'}_{ \quad n\quad,\quad m}=\Phi^{\quad n'\quad, \quad m'}_{5-n-n', 6-m-m'}.
\label{eq:Z3identification}
\eea
having conformal dimension
\bea
\Delta^{n,m}_{n',m'}=\frac{
3\left[6\left(n+n'\right)-5\left(m+m'\right)\right]^2+
\left[6\left(n-n'\right)-5\left(m-m'\right)\right]^2
-12}
{360}
\nonumber
\eea
The fusion algebra between the primary fields is invariant under the
substitution 
\begin{equation}
\Phi\to e^{\frac{2\pi i}{3}q}\Phi,\label{eq:Z3transformation}
\end{equation}
where the pertinent $\mathbb{Z}_{3}$-charge $q$ is defined in
\cite{Fateev1988} as
\bea
q=q^{n,m}_{n',m'}=\left(m-m'\right) \mbox{mod} 3\;.
\label{theZ3charge}
\eea
The three fields in the identification (\ref{eq:Z3identification}) have the same
conformal dimension and $\mathbb{Z}_{3}$ charge.
As a consequence, the OPE coefficients
and the three-point functions vanish unless the total $\mathbb{Z}_{3}$
charge is zero, which severely constrains the possible fusion processes.
Note that the dimension-$3/5$ fields $\Psi,\Psi^*$ are the parafermions $\psi_1,\psi_2$ of \cite{Lukyanov1988}, while the dimension $1/9$ fields are the spin fields $\sigma_1,\sigma_2$.

\begin{table}
	\centering
	\def\arraystretch{1.2}
	\begin{tabular}{c|c|c}
		symbol & $su(3)$ weights notation & $\Delta$ \tabularnewline
		$\mathbb{I}$ & $\Phi^{1,1}_{1,1}=\Phi^{1,1}_{3,4}=\Phi^{3,4}_{1,1}
		$ & $0$ \tabularnewline
		$\varepsilon$ & $
		\Phi^{1,2}_{2,2}=\Phi^{2,2}_{1,2}=\Phi^{2,2}_{2,2}
		$ & $\frac{1}{10}$ \tabularnewline
		$\varepsilon'$ & $
		\Phi^{1,2}_{1,2}=\Phi^{1,2}_{3,2}=\Phi^{3,2}_{1,2}
		$ & $\frac{1}{2}$ \tabularnewline
		$\Psi^{*}$ & $
		\Phi^{1,1}_{2,1}=\Phi^{2,4}_{1,1}=\Phi^{2,1}_{2,4}
		$ & $\frac{3}{5}$ \tabularnewline
		$\Psi$ & $
		\Phi^{2,1}_{1,1}=\Phi^{1,1}_{2,4}=\Phi^{2,4}_{2,1}
		$ & $\frac{3}{5}$ \tabularnewline
		$\Omega$ & $
		\Phi^{2,1}_{2,1}=\Phi^{1,4}_{2,1}=\Phi^{2,1}_{1,4}
		$ & $\frac{8}{5}$ \tabularnewline
		$\zeta$ & $
		\Phi^{1,1}_{3,1}=\Phi^{1,4}_{1,1}=\Phi^{3,1}_{1,4}
		$ & $2$ \tabularnewline
		$\zeta^{*}$ & $
		\Phi^{3,1}_{1,1}=\Phi^{1,1}_{1,4}=\Phi^{1,4}_{3,1}
		$ & $2$ \tabularnewline
		$\xi$ & $
		\Phi^{1,2}_{1,1}=\Phi^{1,1}_{3,3}=\Phi^{3,3}_{1,2}
		$ & $\frac{1}{9}$ \tabularnewline
		$\xi^{*}$ & 
		$\Phi^{1,1}_{1,2}=\Phi^{1,2}_{3,3}=\Phi^{3,3}_{1,1}
		$ & $\frac{1}{9}$ \tabularnewline
		$\eta$ & 
		$\Phi^{1,1}_{1,3}=\Phi^{3,2}_{1,1}=\Phi^{1,3}_{3,2}
		$ & $\frac{7}{9}$ \tabularnewline
		$\eta^{*}$ & 
		$\Phi^{1,3}_{1,1}=\Phi^{1,1}_{3,2}=\Phi^{3,2}_{1,3}
		$ & $\frac{7}{9}$ \tabularnewline
		$\phi$ &
		$\Phi^{2,2}_{1,1}=\Phi^{1,1}_{2,3}=\Phi^{2,3}_{2,2}
		$ & $\frac{2}{45}$ \tabularnewline
		$\phi^{*}$ & 
		$\Phi^{1,1}_{2,2}=\Phi^{2,3}_{1,1}=\Phi^{2,2}_{2,3}
		$ & $\frac{2}{45}$ \tabularnewline
		$\mu$ & 
		$\Phi^{1,2}_{2,1}=\Phi^{2,3}_{1,2}=\Phi^{2,1}_{2,3}
		$ & $\frac{17}{45}$ \tabularnewline
		$\mu^{*}$ & 
		$\Phi^{2,1}_{1,2}=\Phi^{1,2}_{2,3}=\Phi^{2,3}_{2,1}
		$ & $\frac{17}{45}$ \tabularnewline
		$\rho$ & 
		$\Phi^{2,1}_{1,3}=\Phi^{1,3}_{2,2}=\Phi^{2,2}_{2,1}
		$ & $\frac{32}{45}$ \tabularnewline
		$\rho^{*}$ & 
		$\Phi^{1,3}_{2,1}=\Phi^{2,2}_{1,3}=\Phi^{2,1}_{2,2}
		$ & $\frac{32}{45}$ \tabularnewline
		$\nu$ & 
		$\Phi^{1,3}_{1,2}=\Phi^{1,2}_{3,1}=\Phi^{3,1}_{1,3}
		$ & $\frac{13}{9}$ \tabularnewline
		$\nu^{*}$ & 
		$\Phi^{1,2}_{1,3}=\Phi^{3,1}_{1,2}=\Phi^{1,3}_{3,1}
		$ & $\frac{13}{9}$ \tabularnewline
	\end{tabular}\caption{Primary fields of $\mathbb{Z}_{3}^{(5)}$ and corresponding conformal dimension.\label{tab:Primary-fields-Z3} }
\end{table}

\begin{table}
	\centering
	\begin{tabular}{cccc}
		$\varepsilon'\times\Psi=\varepsilon$ & $\varepsilon\times\Psi^{*}=\varepsilon+\varepsilon'$ & $\varepsilon\times\Psi=\varepsilon+\varepsilon'$ 		\tabularnewline
		 $\varepsilon'\times\Psi^{*}=\varepsilon$ 
		&
		 $\Omega\times\varepsilon=\varepsilon+\varepsilon'$ & $\zeta\times\varepsilon'=\varepsilon'$ 		\tabularnewline
		$\zeta\times\varepsilon=\varepsilon$ & $\zeta^{*}\times\varepsilon'=\varepsilon'$
		 &
		  $\zeta^{*}\times\varepsilon=\varepsilon$ 		\tabularnewline $\varepsilon'\times\varepsilon'=\mathbb{I}+2\varepsilon'$ &
		$\zeta\times\Psi=\Omega$ &
		$\zeta^{*}\times\zeta^{*}=\zeta$
		 \tabularnewline
		 $\Psi\times\Psi^{*}=\mathbb{I}+\Omega$ & $\Omega\times\varepsilon'=\varepsilon$ & $\varepsilon\times\varepsilon=\mathbb{I}+2\varepsilon+2\varepsilon'+\Omega+\Psi+\Psi^{*}$ 	
		 	\tabularnewline
		$\zeta\times\Psi^{*}=\Psi$ 
		&		 $\zeta\times\zeta=\zeta^{*}$ &  $\varepsilon\times\varepsilon'=2\varepsilon+\Omega+\Psi+\Psi^{*}$	\tabularnewline
		  $\zeta^{*}\times\Omega=\Psi$ & $\Psi\times\Psi=\Psi^{*}+\zeta$
		 &
		$\zeta\times\Omega=\Psi^{*}$ 		\tabularnewline
		 $\zeta\times\zeta^{*}=\mathbb{I}$ & $\Psi\times\Omega=\Psi$ & $\Omega\times\Omega=\mathbb{I}+\Omega$ 
		\tabularnewline
		 $\Psi^{*}\times\Omega=\Psi^{*}$ &
		$\zeta^{*}\times\Psi=\Psi^{*}$ & $\zeta^{*}\times\Psi^{*}=\Omega$ 	
	\end{tabular}\caption{Fusion rules for neutral fields. \label{tab:Z3-FusionRules}}
\end{table}

\begin{table}
	\centering{}%
	\begin{tabular}{cccc}
		$\xi\times\zeta=\nu$& $\xi\times\zeta^*=\eta$&	$\xi\times\Psi=\phi$& $\xi\times\Psi^*=\mu$	\tabularnewline
		$\xi\times\Omega=\rho$	& $\xi^*\times\nu=\zeta+\varepsilon'$& $\xi^*\times\eta=\zeta+\varepsilon'$&	$\xi^*\times\phi=\Psi+\varepsilon$		\tabularnewline
		$\xi^*\times\mu=\Psi^*+\varepsilon$	& $\xi^*\times\rho=\Omega+\varepsilon$ & $\xi^*\times\xi=\mathbb{I}+\varepsilon'$
	\end{tabular}\caption{Fusion rules involving $\mathbb{Z}_{3}$-charged fields used in (\ref{chiralZ}).\label{tab:Fusion-rules-Charged}}
\end{table}

The modular $\mathcal{S}$ matrix describes the rearranging of the
characters of the theory under modular transformations \cite{di1997conformal}.
Moreover, it determines the fusion rules of conformal primary operators
\cite{Verlinde1988}. Its most general form can be found in \cite{Fateev1990,frenkel1992}.
Here we present, for the sake of clarity, only the case $\mathbb{Z}_{3}^{(5)}$,
which we have tested against the fusion rules of \cite{Totsuka1996,frenkel1992}
and used to generate the others necessary for this paper and to compute the spin conductance.

We first recall a few basic notations from
Lie algebras \cite{di1997conformal}. Denote the $su(3)$ fundamental
weights by $\omega_{1}$ and $\omega_{2}$ and the fundamental roots
by $\alpha_{1}$ and $\alpha_{2}$. The Weyl vector $\rho=\omega_{1}+\omega_{2}$
is the sum of the fundamental weights. A generic weight is then expanded
on this basis as $\omega=\lambda_{1}\omega_{1}+\lambda_{2}\omega_{2}$.
A Weyl reflection of the weight $\omega$ with respect to the root
$\alpha_{j}$ is denoted by $s_{j}\omega$. It is possible to apply
repeatedly a Weyl reflection and construct all the independent strings
$s$ of length $\mbox{len}\left(s\right)$, which constitute the Weyl
group $W$. For $su(3)$, there are $6$ elements:
\begin{center}
	\begin{tabular}{ccc}
		element $s$ & $s\omega$ & signature $\left(-1\right)^{\tiny\mbox{len}(s)}$\tabularnewline
		$\mathbb{I}$ & $\lambda_{1}\omega_{1}+\lambda_{2}\omega_{2}$ & $+1$\tabularnewline
		$s_{1}$ & $-\lambda_{1}\omega_{1}+\left(\lambda_{1}+\lambda_{2}\right)\omega_{2}$ & $-1$\tabularnewline
		$s_{2}$ & $\left(\lambda_{1}+\lambda_{2}\right)\omega_{1}-\lambda_{2}\omega_{2}$ & $-1$\tabularnewline
		$s_{2}s_{1}$ & $\lambda_{2}\omega_{1}-\left(\lambda_{1}+\lambda_{2}\right)\omega_{2}$ & $+1$\tabularnewline
		$s_{1}s_{2}$ & $-\left(\lambda_{1}+\lambda_{2}\right)\omega_{1}+\lambda_{1}\omega_{2}$ & $+1$\tabularnewline
		$s_{1}s_{2}s_{1}=s_{2}s_{1}s_{2}$ & $-\lambda_{2}\omega_{1}-\lambda_{1}\omega_{2}$ & $-1$\tabularnewline
	\end{tabular}
	\par\end{center}
We now define the function
\begin{equation}
\phi_{a}\left(b\right)=\sum_{s\in W}\left(-1\right)^{\tiny\mbox{len}(s)}
e^{-2\pi i\left(a+\rho\right)s\left(b+\rho\right)},\label{phiFn}
\end{equation}
where the sum runs over the elements of the Weyl group.
As a primary field $\Phi^{n,m}_{n',m'}$ is associated with the pair of weights $\lambda=\left(n-1\right)\omega_{1}+\left(n'-1\right)\omega_{2}$
and $\lambda'=\left(m-1\right)\omega_{1}+\left(m'-1\right)\omega_{2}$,
such pairs of weights can be used to label the rows and the columns 
of the modular $\mathcal{S}$ matrix. In this notation, 
\begin{eqnarray}
\mathcal{S}_{(\lambda,\lambda'),(\mu,\mu')} & = & e^{2\pi i\left[\left(\lambda+\rho\right)\cdot\left(\mu'+\rho\right)+\left(\lambda'+\rho\right)\cdot\left(\mu+\rho\right)\right]}\nonumber\\
&&\times
\phi_{\mu+\rho}\left(\frac{6\left(\lambda+\rho\right)}{5}\right)
\phi_{\mu'+\rho}\left(\frac{5\left(\lambda'+\rho\right)}{6}\right). \label{ModularS_Z3}
\end{eqnarray}
The fusion coefficients are finally computed by using the Verlinde
formula \cite{Verlinde1988},
\begin{equation}
\mathcal{N}_{(\lambda,\lambda'),(\mu,\mu')}^{(\nu,\nu')}=\sum_{\left(\eta,\eta'\right)}\frac{\mathcal{S}_{(\lambda,\lambda'),(\eta,\eta')}\mathcal{S}_{(\mu,\mu'),(\eta,\eta')}\mathcal{S}_{(\nu,\nu'),(\eta,\eta')}}{\mathcal{S}_{\mathbb{I},(\eta,\eta')}},\label{Verlinde}
\end{equation}
where the sum runs over the pairs of weight labeling the primary fields,
taking into account the identification (\ref{eq:Z3identification}).

\section{Identity of partition functions}\label{app:Partitionfunction}

One can show the equivalence of partition functions (\ref{eq:ZOOaff}) and (\ref{eq:ZOO}) by using a well-known formula \cite{goddard1986}
for the product of the $su(2)_{k}$ characters of the spin-$l$ primary and $su(2)_{1}$ character of the spin-$l'$ primary,
\begin{equation}
 \chi_{l}^{(k)}\left(q\right)\chi_{l'}^{(1)}\left(q\right)=
 \sum_{j=1}^{k+2}\chi_{\frac{j-1}{2}}^{(k+1)}\left(q\right)\chi_{2l+1,j}^{\mathcal{M}_{k+2,k+3}}\left(q\right),
\end{equation}
where the sum runs over the values $j=(2l+1+2l')~\mbox{mod}~2$.
On the r.h.s. of this equation, we encounter products of a $su(2)_{k}$ character and a character of the 
Virasoro primary labeled by the integers $(2l+1,j)$ in the Kac table of the minimal unitary model $\mathcal{M}_{k,k+1}$.
 Applying twice this relation to the partition functions of three chains with open boundary conditions (\ref{eq:ZOOaff}), one obtains
\begin{eqnarray}
 Z_{OO}^{eee}\left(q\right)&=&\chi_{\mathbb{I}}^{\textrm{\tiny SU(2)}_3}\left(q\right)\chi_{0}^{TI}\left(q\right)\chi_{0}^{I}\left(q\right)
 +\chi_{1}^{\textrm{\tiny SU(2)}_3}\left(q\right)\chi_{3/5}^{TI}\left(q\right)\chi_{0}^{I}\left(q\right)
\nonumber\\
&&
+\chi_{\mathbb{I}}^{\textrm{\tiny SU(2)}_3}\left(q\right)\chi_{3/2}^{TI}\left(q\right)\chi_{1/2}^{I}\left(q\right)
+\chi_{1}^{\textrm{\tiny SU(2)}_3}\left(q\right)\chi_{1/10}^{TI}\left(q\right)\chi_{1/2}^{I}\left(q\right), \nonumber\\
Z_{OO}^{ooo}&=&\chi_{1/2}^{\textrm{\tiny SU(2)}_3}\left(q\right)\chi_{1/10}^{TI}\left(q\right)\chi_{1/2}^{I}\left(q\right)
+\chi_{3/2}^{\textrm{\tiny SU(2)}_3}\left(q\right)\chi_{3/2}^{TI}\left(q\right)\chi_{1/2}^{I}\left(q\right)  \nonumber\\
&&+\chi_{1/2}^{\textrm{\tiny SU(2)}_3}\left(q\right)\chi_{3/5}^{TI}\left(q\right)\chi_{0}^{I}\left(q\right)
+\chi_{3/2}^{\textrm{\tiny SU(2)}_3}\left(q\right)\chi_{0}^{TI}\left(q\right)\chi_{0}^{I}\left(q\right)
\label{eq:ZITI},
\end{eqnarray}
expressed in terms of the characters of the Ising (I) and the Tricritical Ising (TI) models,
here labeled by the dimension of the primary.
The next step is to use the identities \cite{Affleck2001}
\begin{eqnarray}
 \chi_{0}^{I}\left(q\right) \chi_{0}^{TI}\left(q\right)+\chi_{1/2}^{I}\left(q\right) \chi_{3/2}^{TI}\left(q\right)
 &=&\chi^{\mathbb{Z}_3}_{\mathbb{I}}\left(q\right)+\chi^{\mathbb{Z}_3}_{\zeta}\left(q\right)+\chi^{\mathbb{Z}_3}_{\zeta^*}\left(q\right), \nonumber\\
  \chi_{0}^{I}\left(q\right) \chi_{3/5}^{TI}\left(q\right)+\chi_{1/2}^{I}\left(q\right) \chi_{1/10}^{TI}\left(q\right)
 &=&\chi^{\mathbb{Z}_3}_{\Psi}\left(q\right)+\chi^{\mathbb{Z}_3}_{\Psi^*}\left(q\right)+\chi^{\mathbb{Z}_3}_{\Omega}\left(q\right),
\end{eqnarray}
relating the product of characters of the Ising and tricritical Ising models to sums of characters of the $\mathbb{Z}_3^{(5)}$ CFT.
This brings (\ref{eq:ZITI}) into the form (\ref{eq:ZOO}) which is our starting point.

\subsection*{Acknowledgments}

We thank C. Chamon, J. C. Xavier, F. Ravanini and E. Ercolessi for discussions and the High-Performance Computing Center (NPAD) at UFRN for providing computational resources. We acknowledge support by the Deutsche Forschungsgemeinschaft within the network
CRC TR 183 (project C01), by the Alexander von Humboldt Foundation, and by the Brazilian ministries MEC and MCTIC.  \\

\bibliographystyle{elsarticle-num}
\bibliography{thebib}

\end{document}